 \documentstyle[preprint,aps]{revtex}
 \def\half{\mbox{$1 \over 2$}}

 \def\beq{\begin{equation}}
 \def\eeq{\end{equation}}
 \def\beqa{\begin{eqnarray}}
 \def\eeqa{\end{eqnarray}}
 \def\lf{\nonumber \\}
 \def\LP{\left(}
 \def\RP{\vphantom{\half} \right)}
 
 \def\real{\Re {\rm e}}

 \gdef\aver#1{\left\langle #1 \right\rangle}
 \gdef\s#1{\! #1 \!}
 \gdef\l#1{\> #1 \>}
 \gdef\Eq#1{Eq.~(\ref{#1})}
 \gdef\tensor#1{\underline {\underline #1}}
 \gdef\vec#1{{\bf #1}}

\begin{document}

\draft

\preprint{NSF-ITP-95-39}

\title{
Semiclassical Analysis of the Quantum Interference Corrections \\ to the
Conductance of Mesoscopic Systems }

\author{Nathan Argaman\cite{byline}}

\address{
Department of Condensed Matter Physics, The Weizmann Institute of Science, \\
Rehovot 76100, Israel \\ and \\
Institute for Theoretical Physics and Department of Physics, \\
University of California, Santa Barbara, CA 93106
}

\date{May 11, 1995}

\maketitle

\bigskip\bigskip

\centerline{\large Abstract}

\bigskip

%\begin{abstract}

The Kubo formula for the conductance of a mesoscopic system is
analyzed semiclassically, yielding simple expressions for both weak
localization and universal conductance fluctuations.  In contrast to
earlier work which dealt with times shorter than $O(\log \, \hbar^{-1})$,
here longer times are taken to give the dominant contributions.  For
such long times, many distinct classical orbits may obey essentially
the same initial and final conditions on positions and momenta, and
the interference between pairs of such orbits is analyzed.
Application to a chain of $k$ classically ergodic scatterers connected
in series gives $-{1 \over 3} [1-(k \s+ 1)^{-2}]$ for the weak
localization correction to the zero--temperature dimensionless
conductance, and ${2 \over 15} [1-(k \s+ 1)^{-4}]$ for the variance of
its fluctuations.  These results interpolate between the well known
ones of random scattering matrices for $k=1$, and those of the
one--dimensional diffusive wire for $k \rightarrow \infty$.

%\end{abstract}

\bigskip


\section{Introduction}

Semiclassical ideas have been central to the understanding of transport
effects in mesoscopic systems, specifically Weak Localization (WL) and
Universal Conductance Fluctuations (UCF), from the outset
\cite{Khmel,CS,LSF}.  However, they are usually used for handwaving
arguments, with the actual calculation being done by resummation of
perturbation theory expressions \cite{AGD} (Feynman diagrams).  In
recent years, fabrication of ballistic mesoscopic systems has become
feasible, and it has been demonstrated that the chaotic or integrable
nature of the classical dynamics in such systems is reflected in the
quantum interference corrections to their transport properties
\cite{Marcus}.
The perturbative approach with respect to the impurity potential is
inapplicable to such systems, and indeed calculations using other
theoretical tools such as Random Matrix Theory \cite{RMT} (RMT) and
the Nonlinear $\sigma$--Model \cite{Ef} have recently
appeared.  The recent progress in applications of the Semiclassical
Approximation (SCA) to classically chaotic systems \cite{Gutz,LesH}
gives rise to a hope that quantitative results for the mesoscopic
transport effects could be obtained from it.  This hope is realized
below, but it turns out to be necessary to apply the SCA in a somewhat
unorthodox manner.

The reason for the difficulty in the semiclassical description of WL
and UCF is obvious: both effects involve quantum interference
corrections to the classical conductance, which are smaller in powers
of $\hbar$.  Various authors have dealt with this difficulty in
different ways.  Some have followed the diagrammatic derivation quite
closely, and used it for calibration of the magnitude of the effect
\cite{CS}.  Others have limited their attention to effects such as
coherent backscattering, where the quantum interference corrections
appear in the leading order, and are as large as the classical result
\cite{DSF}.  Still others have concentrated on the magnetic field
dependence of UCF, and used RMT to calibrate the magnitude
\cite{JBS,LDJ,HSB}.  All of these studies assume either diffusive or ergodic
classical dynamics, and do not give general semiclassical expressions.

Many of the recent analyses have used the Landauer formula or
scattering approach, rather than the bulk approach of the Kubo
formula.  The proof of the equivalence of these two formulae relies on
the unitarity of the quantum mechanical evolution (current
conservation).  However, semiclassical evolution in classically
chaotic systems is only approximately unitary, and therefore
application of the SCA to the Kubo or Landauer formulae may lead to
different results.  The experience gained from the diagrammatic
calculations \cite{KSL} shows that higher order corrections to the
propagators (diagrams with Hikami boxes \cite{Hikami}) are necessary
for an evaluation of the conductance from the Landauer formula,
but not when the Kubo formula is used.  For this reason, we use the
latter in the present work.

A semiclassical analysis has already been developed for the Kubo
formula in classically chaotic systems by Wilkinson \cite{Wilk}, with
recent applications to transport in antidot arrays \cite{HO_Richter}.
It was assumed there \cite{Wilk} that the relevant propagation times
are shorter than the Ehrenfest time, $t_E \sim O(\log\,\hbar^{-1})$, so
that a coherent state or wave packet maintains its correspondence to
a point in the classical phase space throughout its evolution.
However, it has since become known that semiclassical expressions are
applicable also to later times \cite{OTH}.  At such later times the
region in classical phase space which evolves from an initially
minimal--uncertainty wavepacket is stretched and folded by the
chaotic dynamics into a very long and curved shape, which may
intersect the region defined by some final wave packet several times.
The Ehrenfest time thus marks the onset of interference between
different classical orbits which correspond to these different
intersections (called the mixing regime), and not the breakdown of
the SCA.  In the present work the focus is on times much longer than
the Ehrenfest time, and compact expressions for the contribution of
these interference terms to the conductance are developed.
Interestingly, some of the terms (including the periodic orbit
contributions studied by Wilkinson) retain the same form whether the
orbits under consideration are short or long on the scale of $t_E$,
whereas other terms are peculiar to times longer than $t_E$ and
can not be found by studying the strict semiclassical limit
$\hbar \rightarrow 0$.

The semiclassical expressions derived below exhibit three rather novel
features: firstly, it will be assumed that the actions of the classical
orbits involved are large, and that they thus contribute with essentially
random phases and a statistical description is appropriate; secondly,
interference effects between continuous families of classical paths appear
explicitly;  thirdly, the distribution of classical paths is described as
having a continuous density in phase space, even when the initial position
and momentum are given, i.e.\ it will be necessary to introduce a
small amount of averaging over the initial (and final) conditions.
Such a description is relevant not only for the average over an ensemble
of similar mesoscopic systems which differ in their microscopic
details (the disorder ensemble), but also for a single system at
times longer than the Ehrenfest time.  For example, in the
expression for weak localization, orbits which start at some given position
$\vec r$ and momentum $\vec p$, and end after some time $t$ at the same
position $\vec r$
but at momentum $-\vec p$ will be needed.  Rather than retaining only self
retracing paths, which would be implied by strict classical mechanics, the
results involve the density of orbits in phase space around the point
$(\vec r,-\vec p)$ (note that a self--retracing path is not related to a
distinct path by time reversal, and is thus irrelevant for interference
between {\it pairs} of time reversed paths).  In incorporating these features
we are following the approach of our previous work \cite{AIS}, which pertains
to density--of--states correlations in mesoscopic systems; however the need
to account for contributions which do not strictly obey the initial and final
classical conditions did not arise there.

As in other applications of the semiclassical method, interactions are
ignored (apart from a possibly self--consistent potential), and the
electron fluid is considered as a degenerate thermal distribution of
non--interacting particles moving in a mesoscopic sample.
We have in mind a situation where the electrons' motion is mostly free,
but occasionally they hit an impurity or the boundary of the system and
are scattered.  This corresponds (in the language of quantum chaos
\cite{LesH}) to the motion of a `billiard ball' which is
scattered by some obstacles (see Fig.~1).  In order to apply the SCA,
it is assumed that all the classical dimensions, such as the mean free
path or the radius of curvature of the obstacles, are much larger than
the Fermi wavelength (i.e.\ $\hbar$ is small).  The classical dynamics
is assumed to be completely chaotic, with all orbits hyperbolically
unstable (systems with integrable or intermediate dynamics require
separate consideration).  The number of dimensions or degrees of
freedom, $N$, must thus be larger than $1$.
The exponentially large number of orbits in
such systems justifies our statistical approach.  Possible complications
associated with caustics etc.\ are ignored in the present work, and
only the generic contributions are considered.  In addition we assume
that the spectrum is essentially continuous, i.e.\ that the
single--particle level spacing $\Delta$ is much smaller than the other
energy scales in the system, such as those determined by the
broadening of the levels $\Gamma$ due to the external leads, the
temperature $T$ or the frequency $\omega$ (even when those tend to
zero).  This also implies that there is no exponential Anderson
localization in the system even in two or quasi--one dimensions,
i.e.\ the system is smaller than the localization length, because
$\Delta \ll E_C$ where $E_C$ is the Thouless energy ($E_C=\hbar/t_D$
with $t_D$ the time for an electron to traverse the system).

The various contributions to the conductivity and its fluctuations will be
expressed as integrals over the distribution function of classical orbits
in the system.  Given a Hamiltonian, $H(\vec r,\vec p)$, an initial point
in phase--space, $(\vec r,\vec p)$, and a propagation time $t$, classical
dynamics generates a final point which we denote by $(\vec r_t,\vec p_t)$.
The unaveraged distribution of classical orbits is defined as a $\delta$
function around this point:
\beq
f(\vec r',\vec p',t;\vec r,\vec p) = \delta(\vec r'-\vec r_t) \,
\delta(\vec p'-\vec p_t)  \; .
\eeq
It is convenient to use a distribution function $f_E$ limited to the
energy hypersurface, by factoring out the energy conservation condition:
\beq
f_E(\vec r',\vec p',t;\vec r,\vec p) \,
\delta \! \LP H(\vec r',\vec p') \s- H(\vec r,\vec p) \RP =
f(\vec r',\vec p',t;\vec r,\vec p)  \; .
\eeq
Whereas these distribution functions describe the chaotic classical
dynamics in intricate detail, they can be averaged over small ranges
in initial and final conditions to give a smooth distribution which
describes the evolution in a statistical sense.  This averaging will
be denoted by an overline:
$\overline{ f_E(\vec r',\vec p',t;\vec r,\vec p) }$, with the range of
averaging determined by $\hbar$.  For times $t$
significantly longer than the Ehrenfest time $t_E$, the distribution
$\overline{f_E}$ becomes independent of the specific details of the
averaging procedure.
For example, in an ergodic system $\overline{ f_E }$ becomes
independent of initial and final conditions at long times.
Apart from the distribution function $\overline{f_E}$,
additional properties of the classical paths occasionally appear in the
expressions (e.g.\ the distribution of areas enclosed by the classical
paths is relevant for the case of weak magnetic fields).

Apart from the phase--space averaging denoted by the overline,
angular brackets will be used in order to denote averaging
over the semiclassical phases, which are assumed to be uncorrelated except
for possible symmetries, e.g.\ $\aver{ e^{iS_\alpha / \hbar} } = 0$ and
$\aver{ e^{i(S_\alpha-S_\beta) / \hbar} } = \delta_{\alpha,\beta} +
\delta_{\alpha^T,\beta}$, where $S_\alpha$ and $S_\beta$ are the actions of
classical orbits $\alpha$ and $\beta$, and $\alpha^T$ denotes the orbit
time--reversed to $\alpha$.  This assumes that time--reversal is the
only symmetry in the system, and that the number of orbits is
exponentially large so that the possibility of self--symmetric orbits,
$\alpha = \alpha^T$, may be ignored.
This second averaging should be understood either as an averaging
over the specific positions of the obstacles (the disorder ensemble average),
or, for a single system, as an averaging over a range of possible values of
the Fermi energy $\mu$ (or a range of values of Planck's constant $\hbar$).
The average quantum correction to the conductance will be denoted by
$\aver{\Delta G}$, and its variance by $\aver{ (\Delta G)^2 }_{\text{Var}}$.
The fact that these quantities can be written semiclassically as integrals
over the distribution function $\overline{ f_E }$, without reference to
individual classical orbits and the plethora of their actions and amplitudes,
is very useful in applications to classically chaotic systems \cite{AIS,A}.

The outline of the paper is as follows: in Sec.~II the general
SCA expression for the Kubo conductance is derived, and shown to reduce to
the classical conductance when all interference effects are ignored.
The appropriate choice of the electric field distribution is discussed.
Weak Localization is analyzed in Sec.~III, and Universal
Conductance Fluctuations in Sec.~IV.  In these sections the general
expressions in terms of $\overline{ f_E(\vec r',\vec p',t;\vec r,\vec p) }$
will be applied to diffusive and ergodic systems, for which simple
expressions for $\overline{ f_E }$ are readily available.
We concentrate on calculating the magnitude of the interference
effects at zero temperature and magnetic field, in order not to repeat
the considerations given in the previous semiclassical analyses
\cite{Khmel,CS,DSF,JBS,LDJ}.
Extensions of the analysis, e.g.\ to finite temperatures, and applications to
more complicated systems which consist of several cavities connected
in series through ideal leads, will be considered in Sec.~V, followed by
a discussion in Sec.~VI.  A short description of this work is being
published separately \cite{Aprl}.

\section{The Kubo formula}

According to linear response theory, the real part of the conductivity tensor
is given by (see e.g.\ appendix A of Ref.\ \cite{CS})
\beq
\sigma_{jk}(\omega)  \l=  {e^2 \over \omega m^2} {1 \over {\rm vol}}
\real \sum_{mn} \langle m| \hat p_j |n \rangle
\langle n| \hat p_k |m \rangle
i { f_{FD}(\xi_m)-f_{FD}(\xi_n) \over \hbar \omega + \xi_m - \xi_n +i0}
\; ,
\eeq
where $\xi_m$ and $|m \rangle$ are the single--particle eigenenergies and
eigenstates respectively, $\hat \vec p$ is the momentum operator, and
$f_{FD}(\xi)=1/(\exp[\beta(\xi-\mu)]+1)$ is the Fermi--Dirac distribution,
with $\mu$ the chemical potential and $\beta$ the inverse temperature ($e$
and $m$ are the electron charge and mass respectively, $\omega$ is the
frequency, and ${\rm vol}$ is the volume of the system; we are
assuming a simple effective mass description of the electrons, but the
derivation applies with minor modifications to systems with a
non--spherical Fermi surface).
This can be rewritten in terms of the single--particle propagator [e.g.\
$\delta(\hat H-\epsilon_1) =
\sum_m |m \rangle \delta(\xi_m-\epsilon_1) \langle m|$]
as
\beqa
\sigma_{jk}(\omega)  &\l=&  {e^2 \over m^2} {\hbar \over {\rm vol}}
\real \int d\epsilon_1 \> d\epsilon_2 \>
{i  \over \hbar \omega + \epsilon_1 - \epsilon_2 +i0}
{ f_{FD}(\epsilon_1)-f_{FD}(\epsilon_2) \over \hbar\omega}
\lf && \qquad\quad
\int d\vec r \> d\vec r' \>
\LP {\hbar \over i} {\partial \over \partial r'_j} \>
\langle \vec r'| \delta(\hat H-\epsilon_2) | \vec r \rangle  \RP
\LP {\hbar \over i} {\partial \over \partial r_k} \>
\langle \vec r| \delta(\hat H-\epsilon_1) | \vec r' \rangle  \RP  \; ,
\eeqa
where we have used $\hat p_k =  \int d\vec r \> |\vec r \rangle \,
{\hbar \over i} {\partial \over \partial r_k} \, \langle \vec r|$, and
ignored spin (a spin index summation should be understood with each
spatial integration; we avoid the extra indices below by assuming
spin degeneracy).

For the purpose of a semiclassical analysis, it is convenient to
define quantities which are bilinear in the quantum mechanical
propagators, and to transform them into a form depending on one energy
and one time variable, rather than two energy variables \cite{AIS}.
We are thus led to define the following ``form--factor'':
\beq
\tensor K(E,t;\vec r,\vec r')  \equiv
\int d\epsilon \> e^{-i\epsilon t/\hbar}
\, {\rm Tr} \, \LP \hat \vec j(\vec r') \,
\delta(E + \half \epsilon - \hat H) \, \hat \vec j(\vec r)
\delta(E - \half \epsilon - \hat H) \RP  \; ,
\eeq
where $\hat \vec j(\vec r)$ is the current density operator (for a
given spin projection).  Some of the basic properties of this
form--factor are: (a) it is real; (b) it is symmetric under the
interchange of all indices and the sign of the time variable,
$K_{jk}(E,t;\vec r,\vec r') = K_{kj}(E,-t;\vec r',\vec r)$; and (c)
if an external magnetic field is the only source of time--reversal
symmetry breaking, then reversing the sign of this field (together
with the sign of the spin indices) has the same effect on $\tensor K$
as reversing the sign of $t$.  As we will discuss only weak magnetic
fields and use the SCA, we can rewrite the form--factor as \cite{rems}
\beq \label{K_def}
K_{jk}(E,t;\vec r,\vec r')  \simeq  {e^2 \over m^2} \!
\int \! d\epsilon \, e^{-i\epsilon t/\hbar}
\LP {\hbar \over i} {\partial \over \partial r'_j}
\langle \vec r'| \delta(E+\half \epsilon - \hat H) |\vec r \rangle  \RP \!
\LP {\hbar \over i} {\partial \over \partial r_k}
\langle \vec r| \delta(E-\half \epsilon - \hat H) |\vec r' \rangle \RP
  \, .
\eeq

The conductivity is given in terms of this form--factor as follows:
\beq \label{sigma_FK}
\tensor \sigma (\omega)  \l=  \int dE \> dt \; \real \, F(E,t) \>
\int { d\vec r \> d\vec r'  \over {\rm vol}}\>
\tensor K(E,t;\vec r,\vec r') \;,
\eeq
where
\beq
F(E,t)  \l=
\int {d\epsilon \over 2\pi} \> e^{i\epsilon t/\hbar}
\, {i \over \hbar \omega - \epsilon +i0} \,
{f_{FD}(E \s- \half \epsilon) - f_{FD}(E \s+ \half \epsilon)
\over \hbar\omega}  \; .
%\lf &=&
%{1 \over \hbar\omega} e^{i\omega t} \int d\epsilon
%\int_0^\infty {dT \over \hbar} \> e^{i(\hbar\omega - \epsilon)(T-t)/\hbar}
% \int dE' \> (-f'(E')) \theta(\half \epsilon-|E'-E|)
%\lf & \propto &                          % rewritten as a convolution:
%{i \over \hbar \omega} \int {dT \over \hbar} e^{i\omega T} \theta(T) \;
%{\rm Pr} {\cos(2(E-\mu)(T-t)/\hbar) \over T-t}  \;
%{2(T-t)/\hbar\beta \over \sinh(2(T-t)/\hbar\beta)}  \; .
% \beta is inverse temperature up to factors of 2 and \pi ...
\eeq
%We note that if we take
%$i/(\hbar\omega-\epsilon+i0) \rightarrow \delta(\hbar\omega-\epsilon)$ then
%we find $F(E,t) \rightarrow e^{i\omega t} \, w_1(E) * w_2(E)$ where $w_1$ and
%$w_2$ are two `smearing functions' which are convolved with each other, the
%first corresponding to the temperature (the derivative of $f(E)$) and the
%second corresponding to a box function of width $\omega$.
Causality is reflected by the fact that replacing $\omega$ by
$-\omega$ is tantamount to taking the complex conjugate of $F(E,t)$.
In order to elucidate the structure of $F(E,t)$, note that
$\LP f_{FD}(E \s- \half\epsilon) - f_{FD}(E \s+ \half\epsilon) \RP /\epsilon$
plays a role similar to $\delta(E-\mu)$ (it is a box function of width
$\epsilon$ and height $1/\epsilon$, smeared by the temperature which
will be assumed small; as we will be interested in long times $t$, only small
values of $\epsilon$ are relevant).  In fact, it is straightforward to
show that $\int dE \> F(E,t) =
e^{i\omega t} \theta(t) - {i \over \omega} \delta(t)$, where
$\theta(t)$ is the step function, and the last term cancels with the
diamagnetic term.  The detailed form of $F(E,t)$ becomes considerably
simpler if only the time symmetric combination of its real part is
needed, $\half \real \LP F(E,t)+F(E,-t) \RP$.  Due to the properties
of $\tensor K$ mentioned above, indeed only this combination appears
if one is interested in a symmetric integral over $\tensor K$, such as
the longitudinal conductance, or if the magnetic field vanishes (in
the cases to be considered below both of these conditions are satisfied).
The integration over $\epsilon$ then becomes trivial due to a
$\delta(\epsilon-\hbar\omega)$ factor, and in the limit of
small frequencies we may write
\beq \label{sigma_K}
\tensor \sigma (\omega \rightarrow 0)  \l\simeq
\int_0^\infty dt \int dE \LP -f'_{FD}(E) \RP \>
\int {d\vec r \> d\vec r' \over {\rm vol}} \>
\tensor K(E,t;\vec r,\vec r')  \; .
\eeq
In the zero temperature limit to be considered below, the integration
over $E$ may be omitted, and its value is identified with the
Fermi energy $\mu$.
Notice that the part of $F(E,t)$ which is asymmetric in $t$ and
responsible for the Hall effect has been simplified in this analysis, which
{\it assumes} that $\tensor \sigma$ is a Fermi surface property \cite{BS_S}.

In the case of restricted geometries or non--diffusive systems, the
conductance rather than the volume--averaged conductivity is the
appropriate quantity to study.  One defines the space--dependent
conductivity $\tensor \sigma(\vec r,\vec r')$ exactly as in
Eqs.~(\ref{sigma_FK}) and (\ref{sigma_K}), except that the
integrations over $\vec r$ and $\vec r'$ and the division by the
volume are not performed.  The current is given by
$\vec j(\vec r') = \int d\vec r \, \tensor \sigma(\vec r,\vec r') \,
\vec E(\vec r)$, where
$\vec E(\vec r)$ is the electric field.  The dissipative conductance of a
sample of general geometry with two leads, at zero frequency, can be
written as the dissipated power
$\int d\vec r' \, \vec E(\vec r') \cdot \vec j(\vec r')$,
divided by the voltage $V$ squared:
\beq \label{cdtance}
G = {1 \over V^2} \int d\vec r \, d\vec r' \> \vec E(\vec r') \,
\tensor \sigma(\vec r,\vec r') \,
\vec E(\vec r)  \; .
\eeq
Current conservation is expressed as
$\nabla_\vec r \cdot \tensor \sigma(\vec r,\vec r') =
\nabla_{\vec r'} \cdot \tensor \sigma(\vec r,\vec r') = 0$,
in the absence of a magnetic field
[$\nabla_\vec r \cdot \nabla_{\vec r'} \cdot
\tensor \sigma(\vec r,\vec r') = 0$ still holds even if a magnetic
field is present].  This can be shown by
integration by parts in \Eq{cdtance} (see, e.g., Ref.~\cite{KSL})
to imply that the electric field $\vec E(\vec r)$ need not be
calculated self--consistently, and instead one can use any electric
field distribution which gives the voltage $V$ when integrated along
any path connecting the two leads [the boundary conditions
require that the components of $\vec E(\vec r)$ and of
$\tensor \sigma(\vec r,\vec r')$ perpendicular to an insulating
boundary vanish].  In fact, one may take different electric
field distributions for the two $\vec E(\vec r)$ factors in
\Eq{cdtance} --- in the scattering approach of the Landauer
formula the electric field is concentrated in the source lead for one
factor, and in the drain for the other (another example is the case of
multilead devices, for which the conductance $G_{ij}$ is a matrix, and the
boundary conditions for the two electric field factors may be different).
As already mentioned, the semiclassical expressions derived below for
$\tensor \sigma(\vec r,\vec r')$ do not necessarily obey current
conservation, because higher order corrections in $\hbar$ are not included.
Thus, the SCA expressions for the conductance do depend on the use of
the actual electric field distribution in the sample, as discussed below.

%\section{ The semiclassical theory of the `form--factor' }

The next step is to write down the semiclassical expression for
$\tensor K(E,t;\vec r,\vec r')$ of Eq.~(\ref{K_def}).
As a starting point we use the van Vleck formula:
\beq
\langle \vec r'| \exp(-i \hat H t) |\vec r \rangle  \l=
\sum_{\alpha \in \{ \vec r,\vec r';t \} } A_\alpha \, e^{i S_\alpha / \hbar}
\eeq
where $\alpha$ is a discrete index which runs over all the classical paths
that start at the point $\vec r$ and end after a time $t$ at the point
$\vec r'$.
The amplitude and action of the classical path $\alpha$ are given by
\beq
A_\alpha= i^{\nu_\alpha} \left| \det \LP {1 \over h}
{\partial \vec p_\alpha \over \partial \vec r'} \RP \right|^{1/2}
  \;\;\;\;\; ; \;\;\;\;\;
S_\alpha = \int_\vec r^{\vec r'} (\vec p \> d\vec r - H \> dt)  \; ,
\eeq
where $\vec p_\alpha$ is the initial momentum of the path $\alpha$ and
$\nu_\alpha$ is the integer Maslov index (this and other factors of $i$ may
be ignored for the purposes of the present work).
After Fourier transforming from time $t$ to energy $E$, we have
(see, e.g., \cite{LesH})
\beq \label{SCA}
\langle \vec r'| \delta(E - \hat H) |\vec r \rangle \l=
\sum_{\alpha \in \{ \vec r,\vec r';E \} }
\tilde A_\alpha \> e^{i \tilde S_\alpha/\hbar}  \; ,
\eeq
where the index $\alpha$ counts classical paths of energy $E$, and the
modified amplitude and action are given by
\beq \label{AS}
\tilde A_\alpha  \l=  i^{\tilde \nu_\alpha} \>
\left| \det \LP {1 \over h}
{\partial \vec p_\alpha \over \partial \vec r'} \RP \right|^{1/2} \>
\left| {1 \over h} {dT_\alpha \over dE} \right|^{1/2}
 \;\;\;\;\; ; \;\;\;\;\;
\tilde S_\alpha  \l=  \int_\vec r^{\vec r'} \vec p \, d\vec r  \; .
\eeq
The derivative ${\partial \vec p_\alpha \over \partial \vec r'}$
appearing in this
amplitude is taken at a constant duration of the orbit $t=T_\alpha$ (it is
also possible to re-express this amplitude in terms of an
$(N \s+ 1) \times (N \s+ 1)$ matrix of derivatives taken at constant energy
$E$, but this will not be helpful here; $N$ denotes the number of
dimensions).  Again, the Maslov index $\tilde \nu_\alpha$ may be ignored,
since only the magnitude of $\tilde A_\alpha$ will be needed below.
It is necessary to note the derivatives of the action:
\beq \label{S_der}
{\partial \tilde S_\alpha \over \partial E} = T_\alpha  \;\;\;\; ; \;\;\;\;
{\partial \tilde S_\alpha \over \partial \vec r} = -\vec p_\alpha
\;\;\;\; ; \;\;\;\;
{\partial \tilde S_\alpha \over \partial \vec r'} = \vec p'_\alpha  \; ,
\eeq
where $T_\alpha$ is the duration of the classical path $\alpha$,
$\vec p_\alpha$ is the initial momentum mentioned above, and
$\vec p'_\alpha$ is the final momentum.

Substituting Eq.~(\ref{SCA}) in the expression for the form--factor,
Eq.~(\ref{K_def}), and using the fact that only contributions from
small values of $\epsilon$ will be important (because $t$ is integrated over
a large range) to develop the action around the mean energy $E$, gives
\beqa \label{K_SCA}
\tensor K(E,t;\vec r,\vec r')  &\l\simeq&  {e^2 \over m^2}
\int d\epsilon \> e^{-i\epsilon t/\hbar}
\sum_{\alpha,\beta \in \{ \vec r,\vec r';E \} }
\tilde A_\alpha \tilde A^*_\beta \>
\vec p'_\alpha \, \vec p_\beta \>
e^{i(\tilde S_\alpha+\half \epsilon T_\alpha
      - \tilde S_\beta + \half \epsilon T_\beta)/\hbar}
\lf &=&  {e^2 \over m^2} \,
h \sum_{\alpha,\beta \in \{ \vec r,\vec r';E \} }
\tilde A_\alpha \tilde A^*_\beta \>
\vec p'_\alpha \, \vec p_\beta \>
e^{i(\tilde S_\alpha- \tilde S_\beta)/\hbar} \>
\delta\LP t-{T_\alpha \s+ T_\beta \over 2} \RP  \; .
\eeqa
The $\delta$ function over time should be understood to have a width
determined by the higher order corrections in $\epsilon$.  Inserting this
result in Eqs.~(\ref{sigma_K}) and (\ref{cdtance}) will give the
semiclassical approximation for the conductivity and the conductance.

%\section{Results for the Conductivity}

%\subsection{Drude Conductivity}

Before discussing quantum corrections to the conductivity, we observe how
the classical results can be regained from this expression.
% (e.g., for a diffusive system the Drude conductivity is
% $\sigma_D = {n e^2 \over m} \tau$ where $n$ is the density of electrons
% and $\tau$ is the transport scattering mean free time).
To this end, all interference terms in Eq.~(\ref{K_SCA}) are ignored, and
only the `diagonal' part of the double sum, $\alpha = \beta$, is retained:
\beq \label{K_D}
\tensor K^D(E,t;\vec r,\vec r')  \l=  {e^2 \over m^2} \,
h \sum_{\alpha \in \{ \vec r,\vec r';E \} }
|\tilde A_\alpha|^2 \> \vec p'_\alpha \, \vec p_\alpha \>
\delta\LP t-T_\alpha \RP  \; .
\eeq
In order to proceed, the amplitudes Eq.~(\ref{AS}) should be substituted
here.  As similar expressions will be used below, we note here the general
form of a sum of this type:
\beqa \label{trick}
h \sum_{\alpha \in \{ \vec r,\vec r';E \} } |\tilde A_\alpha|^2 \>  &&
\delta(t \s- T_\alpha) (\dots)_\alpha
 \l=  \sum_{\alpha \in \{ \vec r,\vec r';t \} } |A_\alpha|^2
\delta( E_\alpha \s- E ) (\dots)_\alpha  \lf
 && \l=  \int {d\vec r_0 \, d\vec p_0 \over h^N} \>
\delta\LP H(\vec r_0,\vec p_0) \s- E \RP \>
\delta(\vec r_0 \s- \vec r) \, \delta(\vec r_t \s- \vec r') \>
(\dots)_{ (\vec r_0,\vec p_0) }  \lf
 && \l= {1 \over h^N} \int d\vec p_E \, d\vec p'_E \;
 f_E(\vec r',\vec p',t;\vec r,\vec p) \> (\dots)_{ (\vec r,\vec p) }
\; .
\eeqa
Again, the phase space point $(\vec r_t,\vec p_t)$ is that which
evolves from the initial point $(\vec r_0,\vec p_0)$ by following the
classical dynamics for a time $t$.  In the first equality the factor
$\left| {1 \over h} {dT_\alpha \over dE} \right|$ in $|\tilde A_\alpha|^2$
was used to turn from a fixed energy representation to fixed time.  In the
second equality the sum over $\alpha$ was rewritten as an integral over the
initial coordinate and momentum, using the factor
$|A_\alpha|^2 = \left| \det\LP
 {1 \over h} {\partial \vec p_\alpha \over \partial \vec r'} \RP \right|$
(the integration over the initial position is trivial, and in fact
superfluous at this stage).
The result is an integral over the phase space energy hypersurface,
where the properties of the individual paths which were denoted by the dots
are now identified by the initial coordinate and momentum of each path.
Notice that all the determinants of derivatives which appeared in the
amplitudes have been replaced by integrations over $\delta$ functions
\cite{A}, in such a way that allows for the introduction of
the classical distribution function
$f_E(\vec r',\vec p',t;\vec r,\vec p)$ in the last line,
[integration over the energy surface is denoted by $\int d\vec p_E \dots =
\int d\vec p \, \delta\LP H(\vec r,\vec p) \s- E \RP \dots$].

Applying this trick to the diagonal approximation of the form factor,
\Eq{K_D}, gives the classical contribution to the conductivity of
Eq.~(\ref{sigma_K}):
\beq \label{int_f}
\tensor \sigma^{cl}   \l=
{e^2 \over m^2} {1 \over h^N \, {\rm vol}}
\int_0^\infty dt \> \int d\vec r \, d\vec p_E \, d\vec r' \, d\vec p'_E
\; \vec p \, \vec p' \> f_E(\vec r',\vec p',t;\vec r,\vec p)
\eeq
(at zero frequency and temperature).  This may be rewritten as
\beq \label{sigma_cl}
\tensor \sigma^{cl}   \l=
{e^2 \over m^2} \nu \int_0^\infty dt \> \langle \vec p \, \vec p' \rangle_t
\eeq
where $\nu = {1 \over h^N \, {\rm vol}} \int d\vec r \, d\vec p \>
\delta\LP \mu \s- H(\vec r,\vec p) \RP$
is the density of states (implicitly including the spin summation),
and  the last factor is a momentum correlator:
\beq
\langle \dots \rangle_t  \l= {
\int d\vec r \, d\vec p_E \, d\vec r' \, d\vec p'_E  \> \dots \>
f_E(\vec r',\vec p',t;\vec r,\vec p) \over \int d\vec r \, d\vec p_E }
\eeq
(it is the classical counterpart of the Fermi--surface correlator
$\langle p(0)p(t) \rangle$ defined in \cite{CS}).
Eq.~(\ref{sigma_cl}) is the classical contribution to the Kubo conductivity.
Notice that the density of states contains a factor of $h^{-N}$, so
that the quantum corrections to the conductance, of the order of
$e^2/h$, are small corrections to it (higher powers of $\hbar$).
In the present work, the semiclassical limit is considered with the Fermi
momentum and the mobility (or other characterization of the scattering
potential) taken as classical parameters, so that the density of
electrons and the conductance become trivially $\hbar$ dependent.

For diffusive motion, given some initial value of the momentum $\vec p$, the
average of $\vec p'$ at a time $t$ shortly thereafter is equal to $\vec p$
multiplied by $\exp(-t/\tau)$, where $\tau$ is the momentum relaxation
(or transport) mean free time.  Integrating over the directions of the
initial momentum, the classical conductivity is found to be
diagonal, $\sigma^{cl}_{jk} = \delta_{j,k} \, \sigma^D$, and
\beq
\sigma^{D}(\omega)  \l=  {e^2 \over m^2}
\real \int_0^\infty dt \> \cos(\omega t) \exp(-t/\tau) \, \nu \,
{p_F^2 \over N}
\eeq
where the frequency dependence has been restored ($p_F$ denotes the
Fermi momentum).  Since the density of electrons $n$ is equal to
$\nu {p_F^2 \over 2m} {2 \over N}$, this evaluates to
\beq \label{sigma_D}
\sigma^{D}(\omega)  \l=  {ne^2 \over m} \real {\tau \over 1-i\tau\omega}  \; .
\eeq
which is just the Drude conductivity.

The simple description of the momentum correlations used above is inadequate
for the calculation of the space--dependent conductivity.  In fact, it is
known from diagrammatic theory that when all interference terms are ignored,
it may be written (for $\omega=0$) as \cite{KSL}
\beq \label{sh+long}
\sigma^D_{jk}(\vec r,\vec r')  \l\simeq  \sigma^D(0) \,
[\delta_{j,k} \bar \delta(\vec r \s- \vec r') -
\nabla_j \nabla'_k d(\vec r,\vec r')]  \; .
\eeq
Here $\bar \delta(\vec r \s- \vec r')$ is a smeared $\delta$ function
of range equal to the mean free path $l$, which represents the short
range part of the conductivity (analogous to the Chambers formula),
and is due to paths which have not scattered at all.  The scaled
`diffuson' $d(\vec r,\vec r')$ represents the long range contributions
of paths which have scattered at least once, and obeys the equation
$-\nabla^2 d(\vec r,\vec r') = \bar \delta(\vec r \s- \vec r')$ with
vanishing boundary conditions at the conducting leads and vanishing
normal derivative at insulating boundaries.
This ensures that $\tensor \sigma^D(\vec r,\vec r')$ conserves current.

It is possible to render the long range part unimportant by using the
classical electric field distribution in \Eq{cdtance} (for the standard
rectangular geometry this is just a constant field).  Indeed, if the
condition $\nabla \cdot \vec E = 0$ is valid, then it follows from the
boundary conditions on $\vec E$ and $d(\vec r,\vec r')$ that
$ \int d\vec r \, d\vec r' \> \nabla_j \nabla'_k d(\vec r,\vec r') \,
\vec E(\vec r) \, \vec E(\vec r')$
vanishes by integration by parts.  In this case one may use a simple
approximation to the space--dependent conductivity, keeping only the short
range part [the first term in \Eq{sh+long}], without compromising the
accuracy of the classical conductance $G^{cl}$.
It is emphasized here that this choice of the electric field represents
the actual electric field in the sample, if it is interpreted
as the gradient of the electro--chemical potential, rather than the
externally applied perturbation which can be arbitrary.  In other
words, if charge neutrality is assumed (the chemical
potential can not vary, and the self--consistent electric field is
just that which will not cause any charge perturbations), then the
long range part of the conductivity can not contribute, because it
represents the currents due to the gradient of the induced charge
perturbations.

This idea can be generalized to non-diffusive systems, such as a chaotic
cavity \cite{Levinson} (see Fig.~2).  We will assume that the
classical dynamics in
such a cavity is not only ballistic, but also ergodic, so that the
probability for an electron at any point $\vec r$ inside the cavity
to leave through (or to have come from) the left (right) lead is
proportional to its width $W_L$ ($W_R$).  The self--consistent
electrostatic potential is thus a constant within the cavity, and is
equal to $(W_L V_L + W_R V_R)/(W_L+W_R)$, where $V_L$ and $V_R$
denote the potentials in the corresponding reservoirs.  All of the
potential drop occurs in the leads (or at the boundaries between
the leads and the cavity or the reservoirs).
Note that the velocity correlator of \Eq{sigma_cl} is multiplied
by the electric field in \Eq{cdtance} and integrated over time, in
such a way that the contribution of a certain path [determined by
its initial conditions $(\vec r,\vec p)$] to the conductivity has a
simple interpretation: it is proportional to the potential difference
between the initial point $\vec r$ and the reservoir which an
electron with the given initial conditions will eventually reach.
If the electron passes through an ergodic
cavity on its way, and if the electrostatic potential is chosen
self--consistently, then it is no longer necessary to integrate along
the remainder of the path --- the potential in that cavity is already
the averaged potential of the reservoirs, weighted by the probability
that the electron would leave through the corresponding lead.
The long--range part of the electron paths, i.e.\ following the
electron all the way to the reservoirs, thus becomes unimportant
(in the diffusive case any elastic scattering event which randomizes
the electron's direction of propagation plays the role of an
ergodic cavity, in that it ends the short--range part of the propagation).

According to this discussion, the precise form of the electric field
in the leads does not matter.  For speficity, we take the classical
electric field $\vec E(\vec r)$ to be constant over regions of
size $a$ in each lead.
Combining \Eq{cdtance} with \Eq{int_f}, the classical conductance may be
written as
\beq \label{cl_cdt}
G^{cl}  \l=
{e^2 \over m^2} {1 \over h^N \, V^2}
\int_0^\infty dt \> \int d\vec r \, d\vec p_E \, d\vec r' \, d\vec p'_E \>
\LP \vec p \cdot \vec E(\vec r) \RP \, \LP \vec p' \cdot \vec E(\vec r') \RP
\> f_E(\vec r',\vec p',t;\vec r,\vec p)  \; .
\eeq
The contribution of the short range part to the conductance
thus becomes (the integration over $f_E(\vec r',\vec p',t)$ is trivial):
\beq
G_C^{cl}  \l= {e^2 \over h} \sum_{i=R,L} {\Delta V_i^2 \over V^2 a^2}
\int_i {d\vec r \, d\vec p_E \over h^{N-1}} \> \tau(\vec r,\vec p)
v_F^2 \cos^2(\theta) \; ,
\eeq
where $v_F$ is the Fermi velocity, $\Delta V_i$ denotes the voltage drop over
the corresponding lead, $\theta$ is the angle between $\vec p$ and
the direction of the lead, $\tau(\vec r,\vec p)$ is the time that an
electron starting at $(\vec r,\vec p)$
spends in the region $a$ of the electric field, and $\int_i d\vec r$ denotes
integration over the electric field region in the lead $i$.  The integration
over the directions in $\int d\vec p_E$, together with a factor of
$v_F |\cos(\theta)|$, may be replaced by an integration over the transverse
momentum $\int d\vec p_\perp$ and a summation over the two possible directions
along the lead.  When summed over these two directions, the free time
$\tau(\vec r,\vec p)$ gives $a/v_F/|\cos(\theta)|$.  The
integration over the position along the lead gives a further factor of $a$,
which thus cancels out as it should.  The remaining integral gives the number
of transverse channels in the lead
$g_i = h^{-(N-1)} \int_i d\vec r_\perp \, d\vec p_\perp$, which is
proportional to
the width of the lead (in two dimensions $g_i=2 W_i p_F/h$, in three
dimensions $W_i$ is an area and $g_i=\pi W_i p_F^2/h^2$).  The final result
is (ignoring spin)
\beq \label{cl_fin}
G_C^{cl}  \l=  {e^2 \over h}
\left[ \LP {g_R \over g_R+g_L} \RP^2 g_L +
        \LP {g_L \over g_R+g_L} \RP^2 g_R \right]
\l=  {e^2 \over h} \> {g_R g_L \over g_R+g_L}  \; .
\eeq
This result corresponds to adding the resistances of the two ideal
leads classically in series.  Obviously, in order to reach this result
with any other choice of the electric field factors, one would have to
evaluate also the contributions of the long--range parts of $f_E$ (for
example, the transmission which enters the Landauer formula is due to
paths that start in one lead, scatter inside the cavity, and then leave
through the other lead).

The issue of an appropriate choice of the electric field becomes much more
important for the quantum corrections to the conductivity, to be considered
below.  The reason is that it is quite hard to find a current
conserving approximation for $\tensor \sigma(\vec r,\vec r')$ which
includes these quantum corrections, even in the diffusive case which is
treated well by perturbation theory (see Ref.~\cite{HSB}).  When the
self--consistent electric field configuration is used, current conservation
and the long range part of the conductivity become less important.
For example, if quantum interference gives an enhanced probability to
find an electron at some $(\vec r,\vec p)$ at some time $t$, it
becomes unnecessary to follow the propagation of that electron to possibly
correlated momenta $\vec p'$ at later times \cite{KSLr}.
Loosely speaking, the missing electrons represented by the
non--strictly--vanishing value of $\nabla \cdot \vec j(\vec r)$ may be
thought of as being re--injected into the system with a random
direction, so that the self--consistent potential at $\vec r$
automatically takes care of their contribution.  Unfortunately, the
validity of this approach can only be strictly proven if one can write
down an expression for the higher order corrections to
$\tensor \sigma(\vec r,\vec r')$, and show that they do not contribute
when integrated with the electric field factors.  While this is readily
done in the diagrammatic analysis, it is not easily generalized to
other, non--diffusive systems.  In the following sections we proceed
by analogy with the diffusive case, and calculate the quantum corrections
to the conductivity of a general classically chaotic system, using the
self--consistent electric field configuration and the leading order
quantum corrections to $\tensor \sigma(\vec r,\vec r')$.  We return to
this issue in the final section and show that this approach is indeed
justifiable, at least for an ideally ergodic cavity.

\section{Weak Localization}

In this section the SCA is used to calculate the average of the quantum
correction to the conductivity, i.e.\ the weak localization correction
\cite{CS,Bergmann}.  We concentrate on the long--range part of the
conductivity, i.e. on times $t$ larger than $t_E \sim \tau$ --- the
short--range part does not have a weak localization correction.
As advertised, the actions of the classical
orbits in \Eq{K_SCA} for the form factor $\tensor K(E,t;\vec r,\vec r')$
will be assumed random and uncorrelated, except if the two orbits
are related by a symmetry.  After averaging, only two types of
contributions remain: the classical contribution $\beta=\alpha$, and that
of interference between time reversed orbits $\beta=\alpha^T$ (it is
assumed that time reversal symmetry is the only symmetry in the
system).  For any orbit $\alpha \in \{ \vec r,\vec r';E \}$, we have
$\alpha^T \in \{ \vec r',\vec r;E \}$, with $\alpha$ and $\alpha^T$
sharing the same values of action, amplitude and duration, but
$\vec p_{\alpha^T} = - \vec p'_\alpha$ and
$\vec p'_{\alpha^T} = - \vec p_\alpha$.
The possibility of having a strict equality $\beta=\alpha^T$
arises only in the case that $\vec r'=\vec r$, giving rise to a
factor of 2 enhancement of $\tensor K(E,t;\vec r,\vec r)$ relative to
its classical value.  As the coordinates $\vec r$ and $\vec r'$ are
integrated over, it is necessary to find how this enhancement is
reduced when $\vec r'$ deviates from $\vec r$.
Therefore, we include in the weak localization term all
pairs of orbits for which $\beta \simeq \alpha^T$, in the sense that
$\beta$ and $\alpha$ smoothly deform into a pair of time reversed
orbits when $\vec r'$ approaches $\vec r$.  Orbits which are
self--symmetric are excluded, because their contribution is already
accounted for in the classical term.

The expression for the weak localization correction to the
form--factor thus reads
\beq \label{K_wl}
\aver{ \Delta \tensor K(E,t;\vec r,\vec r') }  \l\simeq
{e^2 \over m^2} \> h
\sum_{\alpha, \beta \in \{ \vec r,\vec r';E \} \atop
\beta \simeq \alpha^T \neq \alpha }
\> |\tilde A_\alpha|^2 \> \delta(t \s- T_\alpha) \>
\vec p'_\alpha (-\vec p'_\alpha) \>
e^{i(\tilde S_\alpha - \tilde S_\beta)/\hbar} \; ,
\eeq
where it is assumed that $\vec r'$ is near $\vec r$, so that the only
important $\vec r'$ dependence is in the rapidly varying phase factor.
A direct evaluation of the $\vec r$ and $\vec r'$ integrations over
the form--factor in the $\hbar \rightarrow 0$ limit gives vanishing
results.  In fact, the stationary phase conditions would imply
$\vec p_\alpha = \vec p_\beta$ and $\vec p'_\alpha = \vec p'_\beta$,
which in classical mechanics can only hold for self--symmetric orbits,
$\alpha = \beta$.  In principle, one could try to evaluate higher order
corrections to the non--stationary phase integrals which arise, in the
$\hbar \rightarrow 0$ limit.  However, in practice
(cf.\ Ref.~\cite{Marcus}) $\hbar$ is not extremely small, and the number
of possible orbits $\alpha$ in a chaotic system can be exponentially
large.  In the mixing regime ($t > t_E$) many of these orbits have such
small momentum differences $\vec p'_\alpha - \vec p'_{\alpha^T}$ so as
to make the phase practically stationary {\it throughout} the
integration region.  The spatial integration region is limited by the
size of the system (in practice it may be smaller because the paths
$\alpha$ and $\beta$ may cease to exist due to caustics or shadowing).
Thus, the contribution of orbits with momentum differences smaller
than $\hbar/l_\perp$ is just proportional to the size of the integration
region, $l_\perp$.  Rather than following the exact distribution of possible
values of $l_\perp$ for different orbits, in the following we describe
this result effectively by a $\delta$ function over the momentum
difference.  Note that this represents a non-standard application of
the SCA, because it is assumed that many orbits can fit into the
width of the $\delta$ function, which is proportional to $\hbar$.
We return to this point in the discussion of Sec.~VI~A.

The weak localization correction to the form factor, \Eq{K_wl}, is
again in the form described by \Eq{trick}, which allows us to
re--express it in terms of the classical distribution of orbits.
Furthermore, \Eq{S_der} may be used to develop the actions around the
point $\vec r'=\vec r$:
\beqa
\tilde S_\alpha(\vec r,\vec r';E)  &\l\simeq&
\tilde S_\alpha(\vec r,\vec r;E) + (\vec r' \s- \vec r) \vec p'_\alpha
\; ; \lf
\tilde S_{\beta}(\vec r,\vec r';E)  &=&
\tilde S_\alpha(\vec r',\vec r;E)  \l\simeq
\tilde S_\alpha(\vec r,\vec r;E) - (\vec r' \s- \vec r) \vec p_\alpha
\eeqa
(inclusion of higher order terms in this expansion turns out to be
unnecessary), giving:
\beq \label{DK_TR}
\aver{ \Delta \tensor K(E,t;\vec r,\vec r') }  \l\simeq
- {e^2 \over m^2} \> {1 \over h^N} \int d\vec p_E \, d\vec p'_E \;
 f_E(\vec r,\vec p',t;\vec r,\vec p) \>
\vec p' \, \vec p' \, e^{i(\vec r'-\vec r)(\vec p+\vec p')/\hbar}  \; .
\eeq
In order to perform the integrations using the stationary phase
approximation, it is necessary to treat the pre--exponential factor in
the integrand as slowly varying.  To this end, we replace
$f_E(\vec r',\vec p',t;\vec r,\vec p)$ with
$\overline{ f_E(\vec r',\vec p',t;\vec r,\vec p) }$, where the overline
denotes averaging of the initial and final positions and momenta over
small ranges.  As long as the range of averaging is much smaller than
a Fermi wavelength in position, and much smaller than a typical value
of $\hbar/l_\perp$ in momentum, there is no way that this replacement can
affect the result of the integrals in \Eq{DK_TR} and \Eq{sigma_K}.
After the replacement, the pre--exponential factor is in fact
smoothly varying, if the time $t$ is indeed significantly longer than
the Ehrenfest time $t_E$ [it is not necessary to explicitly remove
the contributions of self--symmetric orbits from
$\overline{ f_E(\vec r,\vec p',t;\vec r,\vec p) }$ --- such orbits
are exponentially rare in the mixing regime].

The next step uses the stationary phase approximation, specifically
$\int dx \, dp \> f(x,p) \> e^{ixp/\hbar} \simeq h f(0,0)$ for small
$\hbar$ and a smooth $f(x,p)$, in order to perform the integrals over
the angular variables of $\vec p'$ and the transverse components of
$\vec r'$.  The stationary phase conditions identify $\vec p'$ with
$-\vec p$, and the components of $\vec r'$ perpendicular to $\vec p$
with those of $\vec r$.  The longitudinal component of $\vec p'$ is
not integrated over because of the limitation to the Fermi surface,
which gives rise to a factor of $1/v_F$.  The integration over the
longitudinal component of $\vec r'$, parallel to $\vec p$, can not
be done in the stationary phase approximation.  However, as occurs
also in the case of the spectral form--factor \cite{AIS}, this
integration is trivial --- the integrand is constant --- and the
result is just equal to the effective length of the integration region,
which we denote by $l(\vec r,\vec p)$.
This integration region is not limited by a small $\hbar$, and may
extend over relatively long distances along the direction of the
classical path: it represents constructive interference between a
continuous family of classical orbits, labeled by the longitudinal
component of $\vec r'$.  As a result, one has
\beq \label{WL_f}
\aver{ \Delta \tensor \sigma }  \l=
-{e^2 \over m^2} {1 \over h \> {\rm vol}} \int_0^\infty dt
\int d\vec r \, d\vec p_E \; \vec p \, \vec p \>
\overline{ f_E(\vec r,-\vec p,t;\vec r,\vec p) }
\, {l(\vec r,\vec p) \over v_F} \; ,
\eeq
which is a general semiclassical expression for the weak localization
correction to the conductivity [the precise meaning of $l(\vec r, \vec
p)$ will be discussed further below].
Note that the integrations over $\vec r$ and $\vec r'$ always lead to
an identification of $\vec p_\alpha$ with $\vec p_\beta$, and of
$\vec p'_\alpha$ with $\vec p'_\beta$ (cf.\ \cite{rems}).  In the
present case of the weak localization contribution, we also have an
identification of $\vec p$ with $-\vec p'$ (see Fig.~1), leading to
the negative sign of the complete expression.

For diffusive behavior with isotropic scattering, the meaning of
$l(\vec r,\vec p)$ is identified with the free path (see Fig.~1) ---
when $\vec r'$ deviates from $\vec r$ further than the next or
the previous scattering event, the momentum factors in \Eq{K_SCA}
become essentially random (as opposed to the approximation used in
\Eq{K_wl}], leading on the average to a vanishing contribution.
In $N \s= 3$ dimensions, and for times $t > \tau$, one may describe
the distribution of classical orbits by
\beq \label{diff}
\overline{ f_E(\vec r',\vec p',t;\vec r,\vec p) }  \l=
W(\vec r,\vec r';t) {v_F \over 4\pi p^2}  \; ,
\eeq
where $W(\vec r,\vec r';t) \, d\vec r'$ is the probability for a
diffusing particle which started at $\vec r$ to be within $d\vec r'$
of $\vec r'$ at time $t$ (in $N \s= 2$ dimensions $2\pi p$ replaces
$4\pi p^2$ in the last factor).  The factorization of the distribution
into separate spatial and momentum space dependencies, with
$\overline{ f_E }$ independent of the momentum direction, implies that
the factors of $\vec p \, \vec p$ may be replaced by
$\delta_{j,k} p_F^2/N$.  The averaging over the integration segment
$l(\vec r,\vec p)$ gives $2 v_F \tau$ --- the factor of $2$ is due to
$l(\vec r,\vec p)$ being defined as the sum of the free paths in the
backward and the forward directions [another way to justify this factor
is to recall that a classical path with a long integration segment
$l(\vec r,\vec p)$ for the integration over the longitudinal component
of $\vec r'$, will also have the same integration region for that
component of $\vec r$, which means that in the averaging
$l(\vec r,\vec p)$ is weighted by its own length].
The resulting correction to the conductivity is again a diagonal tensor:
\beq \label{sigma_WL}
\aver{ \Delta \sigma_{jk} }  \l=  - e^2 \delta_{j,k} \> {2sD \over h}
\int_\tau^{\tau_\phi}  dt \int {d\vec r \over {\rm vol} } \>
W(\vec r,\vec r;t)  \; ,
\eeq
where $D = v_F^2 \tau /N$ is the diffusion constant, and an $s=2$ spin
degeneracy factor has been explicitly restored.  This result coincides
with Eq.~(3.8) of Chakravarty and Schmid \cite{CS}, and thus their quantum
mechanical derivation of the numerical prefactor in this equation
(their Appendix D, which assumes a white--noise disorder potential)
may be replaced by a semiclassical one.

The time integration appearing here is limited from above by the dephasing
time $\tau_{\phi}$, which is due to interactions of the electron with other
particles not included in the single--particle Hamiltonian (the
assumption of a continuous spectrum implies that the times involved
are shorter than $\hbar / \Delta$).
It is also limited from below, by the mean free time $\tau$.
The probability density $W$ satisfies the diffusion equation:
\beq
{\partial W(\vec r,\vec r';t) \over \partial t} - D \nabla_{\vec r'}^2
W(\vec r,\vec r';t)  \l=
\delta(t) \delta(\vec r-\vec r')  \; .
\eeq
For short times, $W(\vec r,\vec r';t) = (4\pi Dt)^{-N/2} \,
\exp\LP - |\vec r \s- \vec r'|^2/4Dt \RP$.
For times of the order of the diffusion time through the sample $t_D=L^2/D$,
it is necessary to expand $W$ in the eigenfunctions $\Phi_n$ of the diffusion
operator,
$W(\vec r,\vec r';t) = \sum_n \Phi_n(\vec r) \Phi^*_n(\vec r') \exp(-t/T_n)$.
In this representation the integrals over $W(\vec r,\vec r;t)$ in
\Eq{sigma_WL} become trivial, and for $\tau_\phi \rightarrow \infty$,
\beq
\aver{ \Delta \sigma } \l= - {e^2 \over h} {2sD \over {\rm vol} } \sum_n T_n
\; .
\eeq
For example, for a quasi--one--dimensional wire extending from $x=0$
to $x=L$ (with finite cross sectional area, i.e.\ many transverse modes)
the longest lasting eigenmodes are
$\Phi_n(x) \propto \sin(\pi nx/L)$ (where $n=1,2,\dots$) with decay times of
$T_n^{-1} = \pi^2 n^2 D/L^2$.  The boundary conditions are
essential in determining this --- they are closed in the transverse
directions but open in the direction along the wire.  This reflects the fact
that trajectories hitting the latter boundaries will continue through the
hypothesized ideal leads into the reservoirs, and will not return to any
point $\vec r'$ in the sample ($\int W \, dt$ is essentially the
`diffuson' of the diagrammatic technique, which was mentioned
earlier).  Defining the dimensionless conductance per spin direction
$g$, and using the fact that $\sum_n (\pi n)^{-2} = 1/6$, gives
(for the $j=k=1$ component)
\beq \label{G_WL}
\sigma_{11} = s \LP {e^2 \over h} {L^2 \over {\rm vol} } \RP g
\;\;\;\;\; ; \;\;\;\;\;
\aver{ \Delta g } = -{1 \over 3}  \; ,
\eeq
which is a well known result for the diffusive wire geometry.

In order to apply the semiclassical expression, \Eq{WL_f}, to a more
general system, one must specify the meaning of the free path factor
$l(\vec r,\vec p)$, or in other words one must perform the integration
over the longitudinal component of $\vec r'$ with care.
Note that in a two--dimensional system the longitudinal
direction may be defined as the locus of points for which the action
difference $S_\alpha(\vec r,\vec r';E)-S_\beta(\vec r,\vec r';E)$
vanishes --- there is then no reason to stop the integration at the
next scattering event (which in itself may be ill--defined for a
smoothly varying potential).  Following the discussion of the previous
section for the classical conductance of such systems, the value of
the integral is identified as the potential difference
$\Delta V(\vec r,\vec p)$, which is defined as the difference between
the potential at the point eventually reached by an electron at
$(\vec r,\vec p)$ and the point from which it emerged.  Strictly speaking,
$\Delta V(\vec r,\vec p)$ is equal to either $\pm V$ or $0$, because
both the original and the eventual points are in one of the
reservoirs.  However, the quantities in \Eq{WL_f} are averaged, and so
we are led to define $l(\vec r,\vec p) =
\overline{ \Delta V(\vec r,\vec p) } \, p_F /
\LP \vec E(\vec r) \cdot \vec p \RP$
[the averaging in $\overline{ \Delta V(\vec r,\vec p) }$ is over the
same range as in $\overline{ f_E } \,$].
The integration over the longitudinal component of $\vec r'$ is thus
effectively limited by the ``Ehrenfest length'' $v_F t_E$.

With this definition, the application of \Eq{WL_f}
[cf.\ also \Eq{cdtance}] to the chaotic
cavity of Fig.~2, involves $\overline{ \Delta V(\vec r,\vec p) } =
\pm \Delta V_i$, due to the fact that the escape time from the cavity
is assumed to be much longer than $t_E$.  With the specific choice of
the electric field within the leads as before, $|\vec E(\vec r)| =
\Delta V_i/a$, this gives
\beq
\aver{ \Delta G_C }  \l= -{e^2 \over h}  \sum_{i=L,R}
{\Delta V_i^2 \over V^2 a^2} \int_0^\infty dt  \int_i d\vec r \, d\vec p_E \;
v_F \, \cos^2(\theta) \> \overline{ f_E(\vec r,-\vec p,t;\vec r,\vec p) }
\, l(\vec r,\vec p)  \; ,
\eeq
where the factors of $a$ cancel due to the longitudinal component of the
$d\vec r$ integration, and the relation
$l(\vec r,\vec p) |\cos(\theta)| = a$.
The time integral of the distribution function $f_E$ is found by
requiring that
the total number of electrons escaping from the cavity, which can be written
as an integral over any crossection in the leads,
$\sum_{i=R,L} \int_i d\vec r'_\perp \, d\vec p'_\perp \int_{t_E}^\infty dt \>
f_E(\vec r',\vec p',t;\vec r,\vec p)$,
is equal to unity (the escape velocity $v_F |\cos(\theta)|$ is used as before
to transform from $\int_i d\vec p_E \dots$ to $\int_i d\vec p_\perp \dots$).
The time integral $\int \overline{f_E} \, dt$ is independent of the detailed
initial and final positions and momenta due to ergodicity in the cavity, and
is thus equal to $h^{1-N}/(g_L+g_R)$ (as long as $\vec r$ and $\vec r'$
are in the leads, $\vec p$ is in the inward direction, and $\vec p'$
is in the outward direction).
The remaining integral gives just $h^{N-1} g_i$, so that
\beq \label{WL_GC}
\aver{ \Delta G_C }  \l=  -{e^2 \over h}  \left[
\LP {g_R \over g_R+g_L} \RP^2 {g_L \over g_R+g_L} +
\LP {g_L \over g_R+g_L} \RP^2 {g_R \over g_R+g_L} \right]
\l= -{e^2 \over h} \, {g_L g_R \over (g_R+g_L)^2}   \; .
\eeq
In the case of equal leads, $g_L=g_R$, one obtains
$\aver{ \Delta g_C } = -{1 \over 4}$ (with $G=g{e^2 \over h}$), in
agreement with RMT results \cite{RMT}.  Applications to additional
systems will be considered in Sec.~V.

\section{Universal Conductance Fluctuations}

We now turn to the off-diagonal terms $\beta \neq \alpha, \alpha^T$, which
do not contribute to $\tensor \sigma$ on the average, and
calculate the typical magnitude or variance of the fluctuations.
A close inspection of the perturbative derivation (Refs.~\cite{LSF,AS} and
references therein) shows that there are in fact three different types of
contributions.  A sketch of these three types of diagrams and the
corresponding classical paths is given in Fig.~3, and the semiclassical
analysis will be detailed in this section.

The contribution of all $\beta \neq \alpha, \alpha^T$ paths to the
conductivity of Eqs.~(\ref{sigma_K}) and (\ref{K_SCA}) is denoted
here by $\Delta \tensor \sigma^{ND}$.  At first sight it would seem that
each term, defined by a specific choice of $\alpha$ and $\beta$, has an
uncorrelated phase $(S_\alpha \s- S_\beta)/\hbar$, and thus its absolute
magnitude squared gives an independent contribution to
$\aver{ |\Delta \tensor \sigma|^2 }_{\text{Var}}$.  This is indeed true
for terms with $T_\alpha, T_\beta \stackrel{>}{\sim} t_E$, and forms the
first type of contribution [Fig.~3(a) and subsection A below].

If one of $T_\alpha, T_\beta$ is negative [note that according to
\Eq{K_SCA} their sum must be positive], then the integrations over $\vec r$
and $\vec r'$ in \Eq{sigma_K} effectively `join' the paths $\alpha$
and $\beta$ into a single periodic orbit of duration $|T_\alpha|+|T_\beta|$
(the path with negative duration $-|T|$ may be considered as starting at
$\vec r'$ and ending at $\vec r$ after a time $|T|$).
Since a single periodic orbit can be bisected into two segments in
many different ways, all contributing with the same phase, this type of
contribution can show significant interference between different $\alpha,
\beta$ pairs. It is useful to first add up all the contributions
for each periodic orbit, which will be labeled $\gamma$, and then
consider the contribution to the variance from each such term.
This is done below, and forms the second type of contribution
[Fig.~3(b) and subsection B].  As a result of the momentum factors, and
the fact that the integral of the momentum along a periodic orbit must
vanish, these terms do not contribute in the simple cases considered here.

The third type of contribution arises when $T_\alpha$ or $T_\beta$ is
positive, but smaller than the Ehrenfest time.
For definiteness, take $0< T_\beta < t_E$; in this case
the path $\alpha$ forms a periodic orbit, which returns to its
starting coordinate and momentum after a time
$T_\alpha \s- T_\beta$ and then continues along the same direction as
$\beta$ for a time $T_\beta$.  Thus, the action along this last segment
cancels in the expression $(S_\alpha \s- S_\beta)/\hbar$, and just as
in the second case described above, this contribution is close to
a periodic orbit.  These periodic orbit contributions form the third and last
type of contribution to $\aver{ |\Delta \tensor \sigma|^2 }_{\text{Var}}$
[see Fig.~3(c) and subsection C].

In the following subsections, the semiclassical expressions for each of these
three types of contributions are derived.  For simplicity, we consider only
the case for which the temperature and the frequency approach zero, and for
which time reversal symmetry holds.

\subsection{Contributions with
$T_\alpha, T_\beta \stackrel{>}{\sim} t_E$}

The different contributions to
$\aver{ \Delta \sigma^{ND}_{j_1 k_1} \Delta \sigma^{ND*}_{j_2 k_2} }$
will be denoted by $F^1$, $F^2$ and $F^3$, with the $j_1 k_1,j_2 k_2$ indices
suppressed in most of the equations (the complex conjugate of the real
quantity $\Delta \sigma^{ND}_{j_2 k_2}$ is taken for convenience in notation;
it amounts only to exchanging the $\alpha$ and $\beta$ indices).  According
to Eqs.~(\ref{sigma_K}) and (\ref{K_SCA}), the expression for
$|\Delta \tensor \sigma^{ND}|^2$  involves an integration over four different
spatial coordinates.  The first type of contribution, $F^1$
(or at least that part of it that will not vanish after averaging) comes
from regions where both $\vec r$ coordinates and both $\vec r'$
coordinates are close to each other, and both copies of $\alpha$ and of
$\beta$ coincide.  Due to time reversal symmetry, there is also a similar
contribution from the case in which these coordinates are interchanged.
As will become evident, these two cases contribute terms with indices
$\delta_{j_1,j_2} \delta_{k_1,k_2}$ and $\delta_{j_1,k_2} \delta_{k_1,j_2}$
respectively, which are otherwise identical.  The first of these, denoted by
$F^{1a}$ gives
\beqa \label{UCF_B}
&& F^{1a}  =
\LP {e^2 \over m^2} {1 \over {\rm vol}} \RP^2
\int_0^\infty dt_1  \int_0^\infty dt_2
\int d\vec r_+ d\vec r'_+ \int d\vec r_- d\vec r'_-        \lf  && \qquad
h^2 \sum_{\alpha,\beta \in \{ \vec r_+,\vec r'_+;\mu \}}
|\tilde A_\alpha|^2 \, |\tilde A_\beta|^2 \>
\vec p'_\alpha \, \vec p_\beta \, \vec p'_\alpha \, \vec p_\beta \;
\delta\LP t_1-{T_\alpha \s+ T_\beta \over 2} \RP
\delta\LP t_2-{T_\alpha \s+ T_\beta \over 2} \RP \, e^{i \Delta S /\hbar} \,.
\eeqa
The summation here includes only orbits with positive times $T_\alpha$
and $T_\beta$, which scatter at least once, and are not related to each
other by symmetry (other contributions are included in the other terms).
The notation $\vec r_+ = \half(\vec r_1 \s+ \vec r_2)$,
$\vec r_- = \vec r_1 \s- \vec r_2$ etc.\ is used, and only the
contribution for which
$\alpha_1 = \alpha_2$ and $\beta_1=\beta_2$ is retained.  The deviation of
the orbits from $\{ \vec r_+,\vec r'_+,\mu \}$ leads to corrections
to the actions, which to first order are given by
\beq
\Delta S  \l\simeq  - \vec r_- (\vec p_\alpha-\vec p_\beta) +
\vec r'_- (\vec p'_\alpha-\vec p'_\beta)   \; .
\eeq

Consider first the integrations over the time variables.  Necessarily
$t_1 = t_2$, and this time variable will be denoted by $t_+$.
It will be convenient to add a fictitious
integration over $\int dt_- \> \delta\LP t_- - (T_\alpha \s- T_\beta) \RP$.
The $\delta$ functions over time may then be rewritten in the form
$\delta(t_a-T_\alpha) \, \delta(t_b-T_\beta)$, where
$t_a = t_+ \s+ t_- /2$ and $t_b = t_+ \s- t_-/2$.  After this is done, it is
possible to transform the sums over
$\alpha,\beta \in \{ \vec r_+,\vec r'_+;\mu \}$
into phase space integrations using again \Eq{trick}.  This gives
\beqa
F^{1a} & \l= &
\LP {e^2 \over m^2} {1 \over {\rm vol}} \RP^2
\int_{t_E}^\infty  dt_a \int_{t_E}^\infty  dt_b
\int d\vec r_+ d\vec r'_+ \int d\vec r_- d\vec r'_-
{1 \over h^{2N}} \int d\vec p_{a \, E} \, d\vec p'_{a \, E}
\int d\vec p_{b \, E} \, d\vec p'_{b \, E}     \lf  && \qquad\qquad\qquad
 f_E(\vec r'_+,\vec p'_a,t_a;\vec r_+,\vec p_a) \;
 f_E(\vec r'_+,\vec p'_b,t_b;\vec r_+,\vec p_b) \;
\vec p'_a \, \vec p_b \, \vec p'_a \, \vec p_b \; e^{i \Delta S /\hbar}
  \; ,
\eeqa
where the quantities relating to the possible $\alpha$ orbits are
denoted by a subscript $a$, and those of the $\beta$ orbits by $b$.
It is assumed here that the relevant contributions come from orbits
longer than the Ehrenfest time, so that $t_a$ and $t_b$ are bigger
than $t_E$.  This allows us to replace the two factors of $f_E$ by
their smooth averages $\overline{ f_E }$, and to neglect the
contributions of $\alpha = \beta$ to these averaged distributions.
The next step is to use the phase factor in order to perform the
integrations over the relative coordinates and momenta, giving
\beqa \label{F1a}
F^{1a} && \l\simeq  \LP {e^2 \over m^2} {1 \over h \, {\rm vol}} \RP^2
\int_0^\infty  dt_a \int_0^\infty  dt_b  \int d\vec r \, d\vec p_E
\int d\vec r' \, d\vec p'_E
\lf && \qquad\qquad\qquad
\overline{ f_E(\vec r',\vec p',t_a;\vec r,\vec p) } \,
\overline{ f_E(\vec r',\vec p',t_b;\vec r,\vec p) }
\, {l(\vec r,\vec p) \, l(\vec r',\vec p') \over v_F^2} \;
\vec p' \, \vec p \, \vec p' \, \vec p  \; ,
\eeqa
where again the integrations over the longitudinal directions give
factors of the `free path' $l(\vec r,\vec p)$, just as in the evaluation
of the weak localization term.

For diffusive behavior, Eq.~(\ref{diff}) can again be used, and the
momentum direction integrations performed.
The free paths $l(\vec r_0,\vec p_0)$ are replaced as before by
factors of $2 v_F \tau$, giving
\beqa
F^{1a}
%_{j_1 k_1 , j_2 k_2}
&\l=&  s^2 \LP {2e^2 \over h} {L^2 \over {\rm vol}} \RP^2 \>
\delta_{j_1,j_2} \, \delta_{k_1,k_2}              \lf  &&  \qquad
\int_0^\infty {D \, dt_a \over L^2} \int_0^\infty {D \, dt_b \over L^2}
\int d\vec r \, d\vec r' \; W(\vec r,\vec r',t_a) \> W(\vec r,\vec r',t_b)
\eeqa
(again the spin degeneracy factor $s=2$ has been restored).  Using the
decomposition of $W(\vec r,\vec r';t)$ in terms of orthonormal
eigenfunctions, the
spatial integrals give simply $\sum_n \exp(-t_a/T_n) \, \exp(-t_b/T_n)$.
Together with the term $F^{1b}$, this yields
\beq \label{UCF_ND}
%\overline{ \Delta \sigma^{ND}_{j_1 k_1} \Delta \sigma^{ND}_{j_2 k_2} }
F^1  \l=
s^2 \LP {2e^2 \over h} {L^2 \over {\rm vol}} \RP^2 \>
\LP \delta_{j_1,j_2} \delta_{k_1,k_2} + \delta_{j_1,k_2} \delta_{k_1,j_2} \RP
\sum_n \LP {T_n D \over L^2} \RP^2   \; .
\eeq
This expression can be seen to coincide with the first part of Eq.~(46) of
Altshuler and Shklovskii \cite{AS}.  For the example of the quasi--1D wire,
with $T_n D/L^2 = 1/\pi^2 n^2$, this contribution to the conductance
fluctuations gives $\aver{ |\Delta g|^2 }_{\text{Var}} =
{8 \over \pi^4} \sum_{n=1}^\infty n^{-4} = {8 \over 90}$, in terms of
the dimensionless conductance $g$.

A further application of \Eq{F1a} is to the chaotic cavity of Fig.~2.  This
gives (including the factor of 2 due to time reversal symmetry, but ignoring
spin)

\vbox{
\beqa \label{UCF_C}
\aver{ (\Delta G_C)^2 }_{\text{Var}}  & \l= &
2 \LP {e^2 \over h} \RP^2 \sum_{i,j=L,R}
{\Delta V_i^2 \, \Delta V_j^2 \over V^4 a^4}
\int_{t_E}^\infty  dt_a \int_{t_E}^\infty  dt_b
\int_i d\vec r \, d\vec p_E \int_j d\vec r' \, d\vec p'_E
\lf && \qquad\quad
\overline{ f_E(\vec r',\vec p',t_a;\vec r,\vec p) } \,
\overline{ f_E(\vec r',\vec p',t_b;\vec r,\vec p) }
\; l(\vec r,\vec p) \, l(\vec r',\vec p') \; \cos^2(\theta) \,
\cos^2(\theta') \> v_F^2
\lf &=& 2 \LP {e^2 \over h} \RP^2 \sum_{i,j=L,R}
{g_{\bar i}^2 g_{\bar j}^2 \over (g_R+g_L)^4} {g_i \, g_j \over (g_R+g_L)^2}
\lf &=& 2 \LP {e^2 \over h} \RP^2 {g_L^2 \, g_R^2 \over (g_R+g_L)^4}  \; .
\eeqa
}
Here we have used our previous result for the time integral of
$\overline{f_E}$, and all the factors of $a$ have canceled as before.  The
notation $\bar i$ denotes the lead opposite to the lead $i$.  This is just
twice the square of the weak localization result, and in the case of
symmetric leads $g_R = g_L$ reduces to the well known result
$\aver{ (\Delta g_C)^2 }_{\text{Var}} = {2 \over 16}$ (see Ref.~\cite{RMT}).
Notice that in this case all of the conductance fluctuations originate from
the first type of contribution, Fig.~3(a), because there are no periodic
orbits which traverse the region in which the classical electric field does
not vanish.

\subsection{Contributions with $T_\alpha T_\beta < 0$}

Consider next the contributions which are concentrated around periodic orbits.
For definiteness, assume that $T_\beta$ is negative.  The description
$\tilde A^*_\beta e^{-i \tilde S_\beta/\hbar}$ of the path
$\beta \in \{ \vec r,\vec r';E \}$ may then be replaced by the identical term
$\tilde A_{\beta'} e^{i \tilde S_{\beta'}/\hbar}$, associated with
the path $\beta' \in \{ \vec r',\vec r;E \}$ where $\beta$ corresponds
to retracing $\beta'$ backward in time ($T_{\beta'} = -T_\beta >0$).
One may rewrite \Eq{K_SCA} as
\beq
\tensor K(E,t;\vec r,\vec r')  \l\simeq  {e^2 \over m^2} \;
h \sum_{\alpha \in \{ \vec r,\vec r';E \} \atop
\beta' \in \{ \vec r',\vec r;E \} }
\tilde A_\alpha \tilde A_{\beta'} \>
\vec p'_\alpha \, \vec p'_{\beta'} \>
e^{i(\tilde S_\alpha + \tilde S_{\beta'})/\hbar} \>
\delta\LP t-{T_\alpha \s- T_{\beta'} \over 2} \RP  \; ,
\eeq
which is completely equivalent, but more convenient if the times
$T_\alpha$ and $T_{\beta'}$ are positive.  When the spatial
integrations over the form factor are performed, it is seen that
indeed the stationary phase points occur when
$\vec p_{\beta'} = \vec p'_\alpha$ and $\vec p'_{\beta'} = \vec p_\alpha$.
This means that the path $\beta'$ must continue the path $\alpha$, and vice
versa --- together they form a periodic orbit.

The periodic orbits of energy $E$ may be enumerated by the discrete index
$\gamma \in \hbox{$ \{ \vec r \s= \vec r',\vec p \s= \vec p';E \} $}$.
Each periodic orbit is in fact a
continuous family of periodic classical trajectories, which differ from
each other by the choice of the initial position along the orbit.
The contributions of all the different pairs of paths $\alpha$ and $\beta$
which fall along the periodic orbit $\gamma$ must now be found.
Note that if the integral over all positive times $t$ is taken as indicated
in \Eq{sigma_K}, the last $\delta$ function may be replaced by a restriction
to $T_\alpha > T_{\beta'}$.
The next step is to undo the Fourier transforms which led to the energy
representation in terms of $\tilde A_\alpha$ and $\tilde A_{\beta'}$, and to
return to a representation in terms of $A_\alpha$ and $A_{\beta'}$:
\beqa
\int_0^\infty dt \int d\vec r \, d\vec r' \>
\tensor K(E,t;\vec r,\vec r')  & \l\sim &  {e^2 \over m^2} \>
{1 \over h} \int d\vec r \, d\vec r' \> \int_0^\infty dt_1 \int_0^{t_1} dt_2 \;
e^{iEt_1/\hbar} \> e^{iEt_2/\hbar}                  \lf && \qquad\qquad
\sum_{\alpha \in \{ \vec r,\vec r';t_1 \} \atop
\beta' \in \{ \vec r',\vec r;t_2 \} }
A_\alpha A_{\beta'} \> \vec p'_\alpha \, \vec p'_{\beta'} \>
e^{i(S_\alpha + S_{\beta'})/\hbar} \>  \; .
\eeqa
The relation $\sim$ (instead of $\simeq$) is used to indicate that only the
contribution with $T_\beta <0$ is included here.
The contribution of a periodic orbit $\gamma$ may now be calculated.
The spatial integrations here resemble the convolution formula for the
propagator at time $t_1+t_2$ in terms of the propagators at times $t_1$
and $t_2$, of which the trace is then taken.  Thus, apart from the
integration over the time difference $t_1-t_2$ and the appearance of
the momentum factors, the stationary phase integrations can be performed
just as is normally done for the periodic orbits involved in the
Gutzwiller trace formula \cite{Gutz,LesH,A}.  This gives
\beqa \label{periodic}
&& \int_0^\infty dt \int d\vec r \, d\vec r' \> \tensor K(E,t;\vec r,\vec r')
\sim \lf && \qquad\qquad\qquad\qquad\qquad
{e^2 \over m^2}
\sum_{\gamma \in \{ \vec r = \vec r',\vec p = \vec p';E \} }
A^p_\gamma \> e^{i\tilde S_\gamma/\hbar} \>
\int_0^{T_\gamma} dt_0 \int_0^{T_\gamma /2} dt_2 \>
\vec p_\gamma(t_0 \s+ t_1) \, \vec p_\gamma(t_0)  \; ,
\eeqa
where $\vec p_\gamma(t)$ denotes the momentum along the periodic orbit
$\gamma$, at a point parameterized by a time variable $t$, and
$t_1 \s+ t_2 = T_\gamma$.  The periodic orbit amplitude,
\beq
|A^p_\gamma|  \l=  {1 \over h}
\left| \det[M_\gamma-I] \vphantom{\half} \right|^{-1/2}  \; ,
\eeq
where $M_\gamma$ is the monodromy matrix describing the stability of the
orbit $\gamma$, is the same as the Gutzwiller amplitude apart from a time
factor.  As in the case of the non--periodic orbits of the propagator,
the squares of these amplitudes may be written as integrals over the
distribution of classical orbits \cite{AIS},
\beq \label{per_amp}
|A^p_\gamma|^2 \, \delta(t-T_\gamma)  \l=
{1 \over h^2 \, T_\gamma} \int_\gamma dr \, dp_E \;
f_E(\vec r,\vec p,t;\vec r,\vec p)  \; ,
\eeq
where the integration is restricted to the region in phase--space
surrounding the periodic orbit $\gamma$ (the time variable too is
restricted --- the right hand side contains additional contributions
at times $t = m T_\gamma$ with any integer $m$).

\Eq{periodic} implies in fact that these periodic orbit contributions
vanish in all the cases considered in the present work.  One may
extend the time integral of \Eq{sigma_K} to negative times, because
one is interested only in the contribution which is symmetric with respect
to reversal of the magnetic field \cite{BS_S}.  Another way to make
this point is to recall that the time reversed orbit $\gamma^T$ will
also contribute, with $\vec p_{\gamma^T}(t) = - \vec p_\gamma(-t)$.
Thus both time arguments appearing in the momenta $\vec p_\gamma(t)$
can be taken to vary over the whole periodic orbit.  However, the factors of
$\int_0^{T_\gamma} dt \, \vec p_\gamma(t)$ must vanish for a periodic orbit,
in order for it to return to the starting point.
Such a cancelation has also been observed on the diagrammatic level
\cite{AALK}, where the
condition $T_\alpha, T_{\beta'} > 0$ corresponds to the condition that a pair
of propagators has either both propagators advanced or both retarded.
This kind of contribution will not vanish only if the frequency $\omega$ is
non-zero, or if the integral $\int_\gamma \vec E \cdot d \vec r$ along the
periodic orbit is non-zero (this can happen if the electric field is driven
by a time dependent flux, and $\gamma$ surrounds the flux).  These exceptions
will not be discussed further here.

\subsection{Contributions with
$T_\alpha$ or $T_\beta \stackrel{<}{\sim} t_E$}

We now turn to the last type of contribution, which involves cases when
either $|T_\alpha|$ or $|T_\beta|$ is smaller than the Ehrenfest time
$t_E$, so that the stationary phase conditions force the two orbits to
overlap over this period.  In this case too the integrations over the
end points $\vec r$ and $\vec r'$ lead to the appearance of periodic
orbits.
% (strictly speaking, this happens whenever $|T_\alpha|$ or
% $|T_\beta|$ are smaller than the Ehrenfest time, but we only follow
% the contributions with free motion, because the momenta become
% effectively uncorrelated for later times).
In fact, most of the analysis of the previous subsection still holds, but
with $T_{\beta'} <0$: the particle is performing periodic motion along
a path of period $T_\gamma = T_\alpha-T_\beta$ (assuming for the
moment that $T_\beta < t_E$), and after revolving around for a time
$T_\alpha$ it retraces its path for a short period $T_\beta$, and reaches
its starting point.  The corresponding contribution is
\beqa \label{p3}
&& \int_0^\infty dt \int d\vec r \, d\vec r' \> \tensor K(E,t;\vec r,\vec r')
\sim  \lf && \qquad\qquad\qquad\qquad\qquad
{e^2 \over m^2}
\! \sum_{\gamma \in \{ \vec r=\vec r',\vec p = \vec p';E \} } \!\!\!
A^p_\gamma \, e^{i\tilde S_\gamma/\hbar}
\int_0^{T_\gamma} dt_0 \int_{-t_E}^0 dt_2 \,
\vec p_\gamma(t_0 \s- t_2) \, \vec p_\gamma(t_0)  \; .
\eeqa
Note that the only difference between this and \Eq{periodic} is in the
limits of the $t_2$ integration.  The absolute square of each such
term is an independent contribution to the variance of the
conductance, which thus includes a sum over all periodic orbits $\gamma$.

Assuming that the relevant periodic orbits are longer than the
Ehrenfest time, $T_\gamma > t_E$, we identify the $dt_2$ integration
in \Eq{p3} as an integration over the momentum correlator along the
orbit $\gamma$.  A periodic--orbit dependent diffusion
constant $\tensor D_\gamma$ may be defined, such that the time
integrations in Eq.~(\ref{p3}) give simply $m^2 T_\gamma \tensor D_\gamma$.
In order to express the results in terms of the phase--space distribution
function $f_E$, this diffusion coefficient is relabled as
$\tensor D(\vec r,\vec p)$, which is identical to $\tensor D_\gamma$
for all points $(\vec r,\vec p)$ on the
periodic orbit $\gamma$.  Using this notation and the expression for
the amplitudes, \Eq{per_amp}, gives
\beq \label{UCF_3}
F^3  \l=  4 \LP {e^2 \over h \, {\rm vol}} \RP^2 \>
\int_0^\infty dt \> t
\int d\vec r \, d\vec p_E \> f_E(\vec r,\vec p,t;\vec r,\vec p) \>
\tensor D(\vec r,\vec p) \, \tensor D(\vec r,\vec p)  \; .
\eeq
A factor of 2 arises due to contributions
with $T_\alpha < t_E$ rather than $T_\beta < t_E$ [these give the
complex conjugate of \Eq{p3}], and a further factor of 2 allows for time
reversal symmetry, i.e.\ contributions of pairs of orbits with
$\tilde S_\gamma = \tilde S_{\gamma^T}$.

In the case of diffusive motion (with $s=2$), and assuming
$t \gg \tau \sim t_E$,
the diffusion constants $\tensor D_\gamma$ can be approximated by their
average $D \delta_{j,k}$, giving:
\beq \label{F3_diff}
%\overline{ \Delta \sigma^{ND}_{j_1 k_1} \Delta \sigma^{ND}_{j_2 k_2} }
F^3  \l=  \LP {e^2 D \over {\rm vol}} \RP^2 \>
\delta_{j_1,k_1} \, \delta_{j_2,k_2} \> |d_{osc}(\mu)|^2   \; ,
\eeq
where (see Ref.~\cite{AIS})
\beq
|d_{osc}(\mu)|^2  \l= s^2 {4 \over h^2} \int_\tau^{\tau_\phi} dt \;
t \int d\vec r \> W(\vec r,\vec r;t)  \; .
\eeq
In these expressions one sees most explicitly the observation made by
Altshuler and Shklovskii \cite{AS} that the corresponding contribution
to the UCF are associated with the fluctuations in the density of states
($d_{osc}$). The integrations may be done explicitly, giving again a
simple sum over the extinction times of the modes:
\beq \label{UCF_DOS}
F^3  \l=  4 s^2 \LP {e^2 \over h} {L^2 \over {\rm vol}} \RP^2 \>
\delta_{j_1,k_1} \, \delta_{j_2,k_2} \> \sum_n \LP {T_n D \over L^2} \RP^2
  \; .
\eeq
This is identical with \Eq{UCF_ND} apart from the spatial indices, again
in agreement with Eq.~(46) of Altshuler and Shklovskii.  It enhances the
fluctuations of the conductance by 50\%, which thus totals for our example
of the quasi--1D wire to
$\aver{ ( \Delta g )^2 }_{\text{Var}} = {12 \over 90} = {2 \over 15}$,
once again a well known result.

For the calculation of the conductance of chaotic cavities,
\Eq{cdtance}, one must take into account the fact that the electric
field factors are position dependent, and may vary over the
region covered by a periodic orbit $\gamma$.
The periodic--orbit--dependent diffusion constants should be taken
to reflect this:
\beq \label{def_D_gamma}
D_\gamma = {1 \over m^2 \, T_\gamma}
\int_0^{T_\gamma} dt_0 \int_{-t_E}^0 dt_2 \,
\LP \vec p_\gamma(t_0 \s- t_2) \cdot \vec E(\vec r_\gamma(t_0 \s- t_2)) \RP \>
\LP \vec p_\gamma(t_0) \cdot \vec E(\vec r_\gamma(t_0)) \RP  \; .
\eeq
The limit $-t_E$ of the $dt_2$ integration is to be understood in the
same manner as the integration over the longitudinal component of the
$\vec r'$ integrations above --- it is useful to define a free time
$\tau(\vec r,\vec p) = \overline{ \Delta_+ V(\vec r,\vec p) } \, m/
\LP \vec E(\vec r) \cdot \vec p \RP$, in analogy with the free path
$l(\vec r,\vec p)$ defined above.  The potential difference
$\Delta_+ V(\vec r,\vec p)$ is the difference between the reservoir
eventually reached by electrons starting at $(\vec r,\vec p)$ and the
potential at $\vec r$ (the integration is only over the forward
direction, not the backward one \cite{remb}).
Again, although the propagation times to the
reservoirs may be long, the averaging effectively limits the length of
the path contributing to $\tau(\vec r,\vec p)$ by the Ehrenfest length.
Strictly speaking, the averaging in
$\overline{ \Delta_+ V(\vec r,\vec p) }$ should
be taken over the region in phase--space corresponding to a
minimal--uncertainty wavepacket defined by the ``width'' of the
periodic orbit $\gamma$, i.e.\ by the Monodromy matrix $M_\gamma$.
Thus, the fact that the classical path for electrons starting
precisely on a periodic orbit never reaches the reservoirs is
irrelevant.  Our assumption that the Ehrenfest time is much smaller
than the typical propagation times in the system (such as $T_\gamma$
for the relevant orbits) means that
$\overline{ \Delta_+ V(\vec r,\vec p) }$ becomes a smooth function
already with a significantly smaller averaging region, and therefore
the precise form of the averaging is unimportant.

With this definition, \Eq{def_D_gamma} becomes
\beq \label{D_res}
D_\gamma = {1 \over m^2 \, T_\gamma}
\int_0^{T_\gamma} dt_0 \>
\LP \vec p_\gamma(t_0) \cdot \vec E(\vec r_\gamma(t_0)) \RP^2 \>
\tau(\vec r,\vec p) \; .
\eeq
Notice that despite the averaging involved in defining
$\tau(\vec r,\vec p)$, the resulting $D(\vec r,\vec p)$ function can
in principle have large fluctuations, so that it becomes important to
use the actual fluctuating $f_E(\vec r,\vec p,t;\vec r,\vec p)$ in
\Eq{UCF_3}, rather than the smooth averaged one.  A possible way to
avoid this is to use a weighted average for $D(\vec r,\vec p)$, i.e.\
define an averaging such that $\overline{\overline{ D^2(\vec r,\vec p,t) }}
 = \overline{ f_E(\vec r,\vec p,t;\vec r,\vec p) D^2(\vec r,\vec p) }
\> / \> \overline{ f_E(\vec r,\vec p,t;\vec r,\vec p) }$.

As already mentioned, the periodic orbits do not contribute to conductance
fluctuations in the case of the chaotic cavity of Fig.~2 because the electric
field vanishes in the region in which periodic motion can occur,
so that $D(\vec r,\vec p)=0$.  A simple example of a class of chaotic
systems for which the periodic orbits of \Eq{UCF_3} do contribute will be
considered in the next section.  Note that the relationship between
the contribution of periodic orbits to conductance fluctuations and
their contribution to the fluctuations in the density of states is
essentially modified by the presence of the electric field factors,
and no longer follows \Eq{F3_diff}.

\section{Applications to specific systems}

In the previous sections, the semiclassical formulae for weak localization
and universal conductance fluctuations were applied to two simple systems,
the diffusive wire and the ergodic chaotic cavity.  In the case that the
leads are of a constant width, both of the quantum interference effects
are stronger for the diffusive system than for the chaotic scatterer
($-{1\over 3}$ {\it vs}.\ $-{1\over 4}$ for weak localization, ${2\over 15}$
{\it vs}.\ ${2 \over 16}$ for the variance of the conductance).
Furthermore, introducing assymetry in the chaotic cavity by taking
$g_L \neq g_R$ {\it decreases} the interference.  In the present
section systems which are intermediate between these two cases are discussed:
in the first subsection a system of two cavities connected in series is
considered, allowing for any combination of widths of the different leads; in
the second subsection a string of $k$ chaotic cavities in series is
considered, with all leads equal in width.  In the course of the treatment of
these systems, results which are valid more generally, for networks of
ergodic cavities connected by ideal leads, will be given.

Before embarking on the detailed treatment of these special systems, several
possible extensions will be mentioned.  Consider first the effects of
symmetry braking by weak magnetic fields or by spin--flip or spin--orbit
scattering.  This requires knowledge of the area distribution associated with
the paths contributing to
$\overline{ f_E(\vec r',\vec p',t;,\vec r,\vec p) }$, or the
distribution of $2 \times 2$ spin scattering matrices along them, and
has been considered in Ref.~\cite{CS} for weak localization.
Because of the close parralellism
between the present approach and the diagrammatic theory, it is not
surprising that the results for diffusive systems are reproduced.
Specifically, in the case of complete symmetry braking the weak
localization correction either vanishes, if time reversal invariance is
broken, or is multiplied by a factor of $-\half$ for the case of symplectic
symmetry (strong spin--orbit scattering).  The variance of the conductance
is reduced by a factor of 2 or 4, respectively.  These results hold whether
the system is diffusive, ergodic, or simply chaotic.

Temperature and frequency dependences may likewise be treated, and again the
diagrammatic results will be reproduced for the diffusive case.  Observation
of \Eq{K_SCA} shows that the frequency couples to the sum of the periods of
the two interfering orbits $T_\alpha+T_\beta$, whereas temperature (averaging
of $E$ in a range around $\mu$) limits the contributions from orbits with
large period differences $|T_\alpha-T_\beta|$.  This gives different
behaviors for the four types of interference effects considered (weak
localization, and the three different contributions to conductance
fluctuations).
The temperature and magnetic field dependences of UCF, for ergodic cavities
without time--reversal symmetry, were very recently studied by Efetov
\cite{Ef} using the nonlinear $\sigma$--model.  As noted there, the results
are in agreement with previous semiclassical analyses \cite{DSF,JBS,LDJ},
except for the amplitude of the effect which has been reproduced
semiclassically only in the present work.
The most striking result of this reference is that temperature smearing, as
opposed to dephasing, does not change the typical area which enters into the
magnetic field dependence, although it does change the form of that
dependence somewhat.  This is of direct relevance to the experimental work
\cite{Marcus2}, as it may allow the measurement of the dephasing rate
from the magnetic field dependence, by inferring this typical area.
It may be intuitively explained by noting that as the finite temperature
limits only the difference between $T^\alpha$ and $T^\beta$, the areas they
encircle may be arbitrary, and the area difference will be of the same order
as the typical area which enters in the zero temperature case.  Dephasing,
on the other hand, limits the sum of $T^\alpha$ and $T^\beta$, leaving only
the contributions with typically smaller areas and area differences.

\subsection{Two ergodic cavities in series}

The system considered here consists of two ergodic cavities connected in
series through a lead with $g_M$ conducting modes, with $g_L$ and $g_R$
denoting the number of modes in the right and left leads as before.  In
this case ergodicity is not achieved, and there are two different regions
of initial conditions which must be considered in order to evaluate
$\overline{ f_E(\vec r',\vec p',t;,\vec r,\vec p) }$.  For electrons
originating in the left
cavity, the probability of leaving the system through the left lead is
the sum of a geometric series:
$P_{L,L} = {g_L \over g_L + g_M}
\LP 1 - {g_M \over g_L + g_M} {g_M \over g_R + g_M} \RP^{-1}
= {g_L (g_R+g_M) \over g_M g_R + g_R g_L + g_L g_M}$,
whereas the probability of leaving the system through the right lead is
$P_{L,R} = {g_M \over g_L + g_M} {g_R \over g_R + g_M}
\LP 1 - {g_M \over g_L + g_M} {g_M \over g_R + g_M} \RP^{-1}
% = MR / [ L(R+M)+MR ]
= {g_M g_R \over g_M g_R + g_R g_L + g_L g_M} $.
The probabilities for an electron in the right cavity (or situated in one of
the leads but headed towards the right cavity) is given by similar
expressions for $P_{R,L}$ and $P_{R,R}$.  The classical electrostatic
potential in the left cavity is therefore
${g_L (g_R+g_M) V_L + g_M g_R V_R \over g_M g_R + g_R g_L + g_L g_M} $,
and that in the right cavity
${g_R (g_L+g_M) V_R + g_M g_L V_L \over g_M g_R + g_R g_L + g_L g_M} $.
The potential drops in the left, middle and right leads are
\beqa \label{E_dist}
\Delta V_L &\l=& V {g_M g_R \over g_M g_R + g_R g_L + g_L g_M}  \; ; \lf
\Delta V_M &\l=& V {g_R g_L \over g_M g_R + g_R g_L + g_L g_M}  \; ;
\vphantom{ {{\sum \over \sum} \over {\sum \over \sum}} } \lf
\Delta V_R &\l=& V {g_L g_M \over g_M g_R + g_R g_L + g_L g_M}  \; .
\eeqa
The classical conductance for this system may be written as
\beq \label{G2C}
G_{2C}  \l=  {e^2 \over h} \sum_{i=L,M,R} g_i \LP {\Delta V_i \over V}\RP^2
%\left[ g_L \LP {g_M g_R \over g_M g_R + g_R g_L + g_L g_M} \RP^2 +
%g_M \LP {g_R g_L \over g_M g_R + g_R g_L + g_L g_M} \RP^2 +
%\right. \lf && \qquad\qquad\qquad\qquad\qquad\qquad\qquad \left.
%g_R \LP {g_L g_M \over g_M g_R + g_R g_L + g_L g_M} \RP^2 \right]
\l= {e^2 \over h} {g_L g_M g_R \over g_M g_R + g_R g_L + g_L g_M}  \; .
\eeq
This is the result of adding classically in series the resistances
$h/(e^2 g_i)$ of the three leads [cf.\ \Eq{cl_fin} for the case of a
single cavity] .

The weak localization correction for this system is given as a sum of four
terms, corresponding to different $(\vec r,\vec p)$ integration
regions in \Eq{WL_f}:
electrons in the left lead moving towards the left cavity, in the middle
lead moving towards either the left or the right cavity, and in the right
lead moving into the right cavity.  It is convenient to generalize the
probabilities $P_{i,j}$ mentioned above, so as to allow indices which
describe electrons in the middle lead, moving either to the right
$i=M_\rightarrow$, or to the left $i=M_\leftarrow$:
$P_{i,j} = \int dt \int_j d\vec r'_\perp \, d\vec p'_\perp
f_E(\vec r',\vec p',t;\vec r,\vec p)$, where $(\vec r,\vec p)$
can be any phase space point in the `directed lead' $i$, and the time
integration excludes very short times for which no chaotic scattering
has occurred.
The sixteen `probabilities' $P_{i,j}$ can be obtained from the four
probabilities $P_{i,j}$ with $i,j=L,R$ by noting that $i=M_\leftarrow$ is
equivalent to $i=L$, $i=M_\rightarrow$ is equivalent to $i=R$, and similar
equivalences can be obtained for the final condition, $j$, up to factors
of $g_M/g_L$ and $g_M/g_R$, respectively (the directions of the arrows for
$i=L,R$ is obvious and omitted in the notation).  In analogy with \Eq{WL_GC},
one obtains
\beqa \label{WL2C}
\aver{ \Delta G_{2C} }  &=&  -{e^2 \over h}
\sum_{i=L,M_\leftarrow,M_\rightarrow,R}
\LP {\Delta V_i \over V} \RP^2 P_{i,i^T}
%\left[
%\LP {g_M g_R \over g_M g_R + g_R g_L + g_L g_M} \RP^2
%{g_L \over g_L + g_M}
%\LP 1 - {g_M \over g_L + g_M} {g_M \over g_R + g_M} \RP^{-1} +
%\right. \lf && \qquad\qquad \left.
%\LP {g_R g_L \over g_M g_R + g_R g_L + g_L g_M} \RP^2
%{g_M \over g_L + g_M}
%\LP 1 - {g_M \over g_L + g_M} {g_M \over g_R + g_M} \RP^{-1} +
%\right. \lf && \qquad\qquad\quad \left.
%\LP {g_R g_L \over g_M g_R + g_R g_L + g_L g_M} \RP^2
%{g_M \over g_R + g_M}
%\LP 1 - {g_M \over g_L + g_M} {g_M \over g_R + g_M} \RP^{-1} +
%\right. \lf && \qquad\qquad\qquad \left.
%\LP {g_L g_M \over g_M g_R + g_R g_L + g_L g_M} \RP^2
%{g_R \over g_R + g_M}
%\LP 1 - {g_M \over g_L + g_M} {g_M \over g_R + g_M} \RP^{-1}  \right]
%=[(MR)^2 L (R+M) + (RL)^2 M (R+M) + (RL)^2 M (L+M) + (LM)^2 R (L+M)]
%             / (LM+MR+RL)^3
%=MRL [MR(R+M) + RL(R+M) + RL(L+M) + LM(L+M)] / (LM+MR+RL)^3
%=MRL [(M+L)R(R+M) + (R+M)L(L+M)] / (LM+MR+RL)^3
\lf &=&
-{e^2 \over h} \, {g_L g_M g_R (g_L+g_M) (g_M + g_R) (g_R + g_L) \over
(g_L g_M + g_M g_R + g_R g_L)^3 }  \; ,
\eeqa
where the index $i^T$ denotes motion time reversed to that denoted by $i$.
In the case of leads of equal width $g_L=g_M=g_R$, this gives
$\aver{ \Delta g_{2C} } = -8/27$, which is intermediate between the $-1/3$
result for diffusive systems, and the $-1/4$ result for completely random
scattering.

The $F^1$ and $F^3$ types of contributions to the fluctuations of the
conductance may also be evaluated in a quite straightforward manner.
The first gives, as in \Eq{UCF_C}:
\beqa \label{UCF1_2C}
\aver{ (\Delta G_{2C})^2 }_{\text{Var}}^1  &\l=&  2 \LP {e^2 \over h} \RP^2
\sum_{i,j=L,M_\leftarrow,M_\rightarrow,R}
{\Delta V_i^2 \, \Delta V_j^2 \over V^4} {g_i \over g_j} P_{i,j}^2
%\lf
%&=& 2 \LP {e^2 \over h} \RP^2
%\left[
%(MR)^2 (MR)^2 {L/L} {L/L+M}^2 +
%(MR)^2 (LR)^2 {L/M} {M/L+M}^2 +
%(MR)^2 (LR)^2 {L/M} {M/L+M}^2{M/M+R}^2 +
%(MR)^2 (LM)^2 {L/R} {M/L+M}^2{R/M+R}^2 +
%
%(RL)^2 (MR)^2 {M/L} {L/L+M}^2 +
%(RL)^2 (LR)^2 {M/M} {M/L+M}^2 +
%(RL)^2 (LR)^2 {M/M} {M/L+M}^2{M/M+R}^2 +
%(RL)^2 (LM)^2 {M/R} {M/L+M}^2{R/M+R}^2 +
%
%(RL)^2 (MR)^2 {M/L} {M/M+R}^2{L/L+M}^2 +
%(RL)^2 (LR)^2 {M/M} {M/M+R}^2{M/L+M}^2 +
%(RL)^2 (LR)^2 {M/M} {M/M+R}^2 +
%(RL)^2 (LM)^2 {M/R} {R/M+R}^2 +
%
%(LM)^2 (MR)^2 {R/L} {M/M+R}^2{L/L+M}^2 +
%(LM)^2 (LR)^2 {R/M} {M/M+R}^2{M/L+M}^2 +
%(LM)^2 (LR)^2 {R/M} {M/M+R}^2 +
%(LM)^2 (LM)^2 {R/R} {R/M+R}^2
%\right]
%{1 \over (g_M g_R + g_R g_L + g_L g_M)^4}
%\LP {(g_L + g_M) (g_M + g_R) \over g_L g_M + g_M g_R + g_R g_L} \RP^2
%\lf
%&=& 2 \LP {e^2 \over h} \RP^2
%L^2 M^2 R^2 \left[
% (MR)^2 (R+M)^2 + R^2 L M (R+M)^2 + R^2 L M M^2 + M^2 L M^2 R +
% R^2 M L (R+M)^2 + (LR)^2 (R+M)^2 + (LR)^2 M^2 + (LM)^2 M R +
% (MR)^2 M L + (LR)^2 M^2 + (LR)^2 (L+M)^2 + L^2 M R (L+M)^2 +
% M^2 R M^2 L + L^2 R M M^2 + L^2 R M (L+M)^2 + (LM)^2 (L+M)^2
%\right]
%{1 \over (g_M g_R + g_R g_L + g_L g_M)^6}
%\lf
%&=& 2 \LP {e^2 \over h} \RP^2
%(MRL)^2 \left[ R^2 (M+L)^2 (R+M)^2 + L R (L+M) (R+M) M^2 +
% R L (L+M) (R+M) M^2 + L^2 (R+M)^2 (L+M)^2 \right]
%{1 \over (g_M g_R + g_R g_L + g_L g_M)^6}
\lf
&=& 2 \LP {e^2 \over h} \RP^2   g_L^2 g_M^2 g_R^2 (g_L + g_M) (g_R + g_M)
\lf && \qquad\qquad\qquad\qquad \times \>
{ (g_L+g_M) (g_R+g_M) (g_L^2+g_R^2) + 2 g_L g_R g_M^2
\over (g_M g_R + g_R g_L + g_L g_M)^6 }
  \; .
\eeqa
Again, in the case $g_L=g_M=g_R$ we find
$\aver{(\Delta g_{2C})^2}^1 = {80 \over 729}$, which is
intermediate between the $2/16$ and $4/45$ results of chaotic and
diffusive systems respectively.
% 2/16 > 16*5 / 3^6 > 4/45

In order to find the contribution of periodic orbits from \Eq{UCF_3},
consider a periodic orbit which traverses the middle lead $n_\gamma$
times, which must be even.  The integral of \Eq{D_res} gives
$D_\gamma = (n_\gamma / T_\gamma) (\Delta V_M^2 /2)$ [the one half
comes from $\tau(\vec r,\vec p)$, which is on the average
$a/ 2 |\cos(\theta)| v_F$ for $\vec r$ in the electric field region].
\Eq{UCF_3} then gives
\beq
\aver{ (\Delta G_{2C})^2 }_{\text{Var}}^3 \l=  4 \LP {e^2 \over h} \RP^2 \>
\sum_{i=M_\leftarrow,M_\rightarrow} {\Delta V_i^4 \over V^4}
\int_0^\infty dt  \int_i d\vec r_\perp \, d\vec p_\perp \;
f_E(\vec r,\vec p,t;\vec r,\vec p) \, {n(\vec r,\vec p,t) \over 4}  \; ,
\eeq
where we have replaced the integration over the whole length of the
periodic orbits, $\int_0^\infty dt \int d\vec r \, d\vec p_E \;
f_E(\vec r,\vec p,t;\vec r,\vec p) \LP n(\vec r,\vec p,t)^2/4t \RP$
by an equivalent term which integrates only over the cross--section of
the lead, $\int_0^\infty dt \sum_i \int_i d\vec r_\perp \, d\vec p_\perp \;
f_E(\vec r,\vec p,t;\vec r,\vec p) \LP n(\vec r,\vec p,t)/4 \RP$.
The notation $n(\vec r,\vec p,t)$ identifies the value of $n_\gamma$
corresponding to the different periodic orbits, as was done with
$D(\vec r, \vec p)$ above \cite{repeti}.

Due to the appearance of the $n(\vec r,\vec p,t)$ factors, this result
cannot be written directly in terms of the `probabilities'
$P_{i,j}$, and requires instead the following modification:
the geometric sum in $P_{M_\leftarrow,M_\leftarrow}$ and
$P_{M_\rightarrow,M_\rightarrow}$ is weighted by a factor of $m/2$ for the
$m$th term in the sum ($n_\gamma=2m$), and thus gives $\half z/(1-z)^2$
where $z=g_M^2/(g_L \s+ g_M)(g_M \s+ g_R)$.  Taking into account the
contributions from both $M_\leftarrow$ and $M_\rightarrow$, one has
\beq \label{UCF3_2C}
\aver{ (\Delta G_{2C})^2 }_{\text{Var}}^3 \l=  4 \LP {e^2 \over h} \RP^2 \>
\LP {g_L g_R \over g_L g_M + g_M g_R + g_R g_L} \RP^4 {z \over (1-z)^2}
\; ,
\eeq
which for $g_L=g_M=g_R$ (with $z=1/4$) gives
$\aver{ (\Delta g_{2C})^2 }^3 = {16 \over 729}$.
This too is intermediate between the zero result for the single
chaotic cavity, and the ${2 \over 45}$ result of the diffusive wire.

Adding the results of Eqs.~(\ref{UCF1_2C}) and (\ref{UCF3_2C}) gives for the
total conductance fluctuations
\beqa \label{UCF_2C}
\aver{ (\Delta G_{2C})^2 }_{\text{Var}} &\l=&
2 \LP {e^2 \over h} \RP^2   g_L^2 g_M^2 g_R^2 (g_L+g_M) (g_R+g_M) (g_L+g_R)
\lf && \qquad\qquad\qquad\qquad \times \>
{ (g_L+g_M) (g_R+g_M) (g_L+g_R) - 2 g_L g_R g_M
\over (g_M g_R + g_R g_L + g_L g_M)^6 }
  \; .
\eeqa
For the case of equal leads this gives
$\aver{ (\Delta g_{2C})^2 }_{\text{Var}} = \LP {2 \over 3} \RP^5$, which is
again intermediate between the ${2 \over 16}$ and ${2 \over 15}$ results.
%(${32 \over 256} < {32 \over 243} < {32 \over 240}$).

The quantum interference effects obtained in Eqs.~(\ref{WL2C}) and
(\ref{UCF_2C}) can be considered in the two limiting cases of a very wide or
very narrow middle lead.  In the case $g_M \gg g_L , g_R$, the middle lead
connects the two cavities very efficiently, so that a particle in one of them
will explore the phase--space of both cavities ergodically before having a
chance to leave through one of the external leads.  Thus, the results of a
single cavity, Eqs.~(\ref{WL_GC}) and (\ref{UCF_C}), are obtained in this
limit.  This can be understood already from the electric field distribution,
\Eq{E_dist}, which vanishes in the middle lead in this case.
In the opposite limit $g_M \ll g_L , g_R$ the middle lead acts as a weak link
(the use of semiclassics assumes $g_i \gg 1$, so it is still much larger than
a quantum point contact), and the quantum interference effects are
suppressed, giving
$\aver{ \Delta G_{2C} } \simeq -(e^2/h) g_M (g_L^{-1}+g_R^{-1})$ and
$\aver{ (\Delta G_{2C})^2 }_{\text{Var}} \simeq
2 (e^2/h)^2 g_M^2 (g_L^{-1}+g_R^{-1})^2$.
As the expressions of Eqs.~(\ref{WL2C}) and (\ref{UCF_2C}) are
symmetric with respect to any permutation of the indices $L,M,R$, it
does not matter whether the middle lead or any one of the other leads
is taken as very wide or very narrow.  However, the ratio of the
periodic orbit contribution, \Eq{UCF1_2C}, to that of interference
between pairs of different paths, \Eq{UCF3_2C}, is not invariant under
such permutations.  In fact, this ratio is maximal when $g_M \ll g_L = g_R$,
and in that case it is equal to 1, which is {\it larger} than the ratio
of $1/2$ familiar from diffusive systems (it is $0$ for a single cavity).
The quantum interference effects turn out to be largest when the three
leads are of equal width, which is the case to be studied and
generalized in the next subsection.

\subsection{Chains of ergodic cavities}

It is clear that by adding more and more ergodic cavities in series, the
situation of the diffusive one--dimensional wire may be approached.  Consider
the case of $k$ cavities connected by leads of equal width, $g_i=const$.
Even though the leads may be identical to each other, the chaotic
cavities are taken to be different in order to avoid a periodic situation in
which Bloch states would emerge (i.e.\ we assume the absence of translation
symmetry).  In this case the electric field configuration is trivial, with
$\Delta V_i/V = 1/(k \s+ 1)$ in all leads.  In order to make use of the
semiclassical formulae of the type appearing in Eqs.~(\ref{WL2C}),
(\ref{UCF1_2C}) and (\ref{UCF3_2C}), one needs to study and generalize
the classical probabilities $P_{i,j}$.

It is convenient to define a classical dynamic probability $p_{l,m}(t)$ equal
to the probability that an electron will be found in the $m$th cavity, given
that it was initially in the $l$th cavity and that it has since traversed
through a lead $t$ times ($l,m=1,\dots,k$ and $t=0,1,\dots$).  Notice that
the `time' variable $t$ is discretized, with no reference to the actual time
the electron may spend in the leads and in the cavities on its way.  Quite
generally, $P_{i,j} = {g_j \over g_m} \sum_{t=0}^\infty p_{l,m}(t)$,
where $l$ is the cavity that the `directed lead' $i$ is flowing into, $m$ is
the cavity out of which $j$ is flowing, and $g_m$ is the sum of the
conductances of the two leads connected to the cavity $m$.
The simple electric field configuration mentioned above is directly
related to these `probabilities'.

A dynamic difference equation, similar to the diffusion equation,
may be written for $p_{l,m}(t)$:
\beq
p_{l,m}(t+1) = \half \, [ \, p_{l,m-1}(t) + p_{l,m+1}(t) \, ]  \; ,
\eeq
with the initial condition $p_{l,m}(0)=\delta_{l,m}$ and the
boundary conditions $p_{l,0}(t) = p_{l,k+1}(t) =0$.  The classical evolution
can be decomposed into $i=1,\dots,k$ eigenvalues $\alpha_i$ and
eigenfunctions $\beta_{i,l}$, such that
$p_{l,m}(t) = \sum_{i=1}^k \beta_{i,l} {\alpha_i}^t \beta_{i,m}$.
For the simple system considered here explicit expressions are available:
$\alpha_i = \cos {i \over k+1} \pi$, and
$\beta_{i,l} = \sqrt{2 \over k+1} \sin {i l \over k+1} \pi$.
Just as has happened for the diffusive case, the eigenfunctions turn out to
be unimportant, and only the eigenvalues $-1 < \alpha_i < 1$ will play a
role.

Generalizing the results of Eqs.~(\ref{WL2C}), (\ref{UCF1_2C}) and
(\ref{UCF3_2C}) to the present case, and rewriting them in terms of the
dynamical probabilities, gives for the weak localization correction
\beq \label{cross_WL}
\aver{ \Delta G_{kC} }
= - {e^2 \over h} {1 \over (k+1)^2} \sum_{l,t} p_{l,l}(t)
= - {e^2 \over h} {1 \over (k+1)^2} \sum_i {1 \over 1- \alpha_i}
= - {e^2 \over h} {1 \over 3} \LP 1 - {1 \over (k+1)^2} \RP  \; ,
\eeq
where the explicit values of $\alpha_i$ were used in the last equality.  The
result for the conductance fluctuations is
\beqa \label{cross_UCF}
\aver{ (\Delta G_{kC})^2 }
& = & 2 \LP {e^2 \over h} \RP^2 {1 \over (k \s+ 1)^4}
\sum_{l,m,t,t'} p_{l,m}(t) \, p_{l,m}(t') +
\LP {e^2 \over h} \RP^2 {1 \over (k \s+ 1)^4}
\sum_{l,t} t \, p_{l,l}(t)
\lf &=& \LP {e^2 \over h} \RP^2 {1 \over (k \s+ 1)^4}
\left( 2 \sum_i {1 \over (1- \alpha_i)^2} +
\sum_i {\alpha_i \over (1- \alpha_i)^2} \right)
\lf &=& \LP {e^2 \over h} \RP^2 {1 \over (k \s+ 1)^4} \, {2 \over 45}
\LP [(k \s+ 1)^2 \s- 1] [2(k \s+ 1)^2 \s+ 7] +
             [(k \s+ 1)^2 \s- 1] [(k \s+ 1)^2 \s- 4] \RP
\lf &=& \LP {e^2 \over h} \RP^2 {2 \over 15}
\LP 1 - {1 \over (k \s+ 1)^4} \RP
\; ,
\eeqa
where the two different contributions are kept separate until the last
equality.  Notice that the factor of $\half = {g_j \over g_M}$ in the
relationship between $P_{i,j}$ and $p_{l,m}$ exactly compensates the fact
that the number of terms in the $i,j$ summations is twice larger than in the
$l,m$ summations (there are twice as many directed leads as cavities).

Naturally, these results reproduce those described above for $k \s= 1$
and $k \s= 2$ cavities ($k \s= 0$ describes an ideal wire), and approach the
diffusive results for $k \rightarrow \infty$.  A similar crossover behavior
has been calculated for a different kind of system which interpolates
between the same ergodic and diffusive limits, using supersymmetry
techniques \cite{IWZ}.  In that case too the idea was to connect $k$ ideally
chaotic cavities in series, but the connections were made directly in the
Hamiltonian, by introducing matrix elements which mix between the quantum
states of adjacent cavities (each cavity was ascribed a GOE Hamiltonian).
The results were as here rational functions of $k$ which reproduce the
$k\s=1$ and the $k \rightarrow \infty$ limits, but their form was much more
complicated than Eqs.~(\ref{cross_WL}) and (\ref{cross_UCF}).  Even the
asymptotic behavior for large $k$ is not similar, and in fact the
crossover for UCF described in Ref.~\cite{IWZ} is slightly non--monotonous.
It would be very interesting to compare our results also with a
continuous crossover between an ideal lead ($k\s=0$ here) and a diffusive
wire, obtained when a disordered region of size $L$ and mean--free--path $l$
is introduced in the lead (with the continuous parameter $L/l$).  In
principle such an analysis can be carried out using the semiclassical methods
developed here, but would depend on the details of the disordered wire
($\overline{ f_E }$ and $\vec E(\vec r)$ depend on whether the
scattering is isotropic or small--angle scattering).

\section{Discussion}

The main results of the present work are the semiclassical expressions
for the mean and variance of the quantum interference corrections to
the conductance, i.e.\ weak localization and universal conductance
fluctuations.  For classically chaotic systems, these
quantities are expressed as integrals over the distribution of
classical orbits, $\overline{ f_E(\vec r',\vec p',t;\vec r,\vec p) }$,
involving also additional quantities which can be derived from this
distribution: the self--consistent electric field $\vec E(\vec r)$, and
the effective free paths $l(\vec r,\vec p)$ and diffusion constants
$D(\vec r,\vec p)$.  Knowledge of this distribution of classical
orbits is readily available in the applications considered here
(see Sec.~V),
due to the assumption of a {\it diffusive} system or a system
consisting of several {\it ergodic} cavities connected through ideal
leads.  The important task of demonstrating the use of these
expressions on a generic system lies beyond the scope of the present
work.  However, it is stressed that finding this classical
distribution numerically for a given potential should be relatively
easy, because only a statistical knowledge of the classical orbits is
necessary, and there is no need to form a full database consisting of
exponentially many orbits.

It was originally thought \cite{Khmel} that the SCA could help to
bridge the gap between the down--to--earth experimentalists and the
abstract mathematical analysis of the theorists.  Unfortunately, the
recent developments in the theory have considerably widened this gap,
with the introduction of diffusion equations in $n$--dimensional
eigenvalue spaces, and supersymmetric techniques.  It is hoped that
the present method will contribute to reversing this trend
\cite{comp}, although it is acknowledged that the semiclassical
approach has its own limitations (in particular, the case of a few,
or partially open, channels is outside its scope).  The following two
subsections give a detailed discussion of the novel features of the
present work, and in that context an attempt is made to bridge a
different gap --- that between the theory of disordered systems and
the theory of quantum chaos.  Possibilities for cross--fertilization
between these two fields, in both directions, are pointed out.

\subsection{Modified Semiclassical Approximation for the Mixing Regime}

It should be emphasized that the semiclassical analysis was applied here in
an unorthodox manner.  A strict stationary phase argument would
require the initial momenta (and final momenta) of the two classical paths
$\alpha$ and $\beta$ appearing in the semiclassical expression for the
conductivity, \Eq{K_SCA}, to be strictly equal to each other.  As these two
paths start at the same position, they would have to be either identical to
each other (the classical contribution to the conductivity), or to differ by
completing a different number of revolutions around a strictly periodic orbit
\cite{Wilk} [as in $F^3$ of \Eq{UCF_3}].  Thus no weak localization
corrections, and no contributions of the first kind [$F^1$ of \Eq{UCF_B}]
to universal conductance fluctuations would be obtained.  However, the
value of $\hbar$ is never infinitesimally small, and thus the stationary
phase argument is never completely strict.  In fact, in chaotic systems
interesting contributions to physical quantities often arise from the
mixing regime, i.e.\ from orbits with propagation times between the
Ehrenfest time $t_E$ and the typical escape time $t_{esc}$.  This regime
disappears in the extreme $\hbar \rightarrow 0$ limit, and
it is thus not surprising that its detailed treatment deviates from
the standard SCA.

Consider for example the classical orbit of Fig.~1, and its behavior when
$\vec r'$ is not strictly equal to $\vec r$.  Labeling the impurities
by digits, we refer to the interference term between the path
$\alpha = \vec r$-1-2-3-4-2-1-$\vec r'$ and the path
$\beta = \vec r$-1-2-4-3-2-1-$\vec r'$, in
which the order of scattering has been reversed.  Clearly $\tilde S_\alpha$
and $\tilde S_\beta$ are guaranteed to be equal only if $\vec r=\vec r'$.
It is also clear [from \Eq{S_der}] that when $\vec r'$ deviates from
$\vec r$, the difference of actions will be proportional to the
difference in the final momenta of these two paths, which for the case
depicted in the figure is about 1\% of $p_F$.  This momentum
difference, $\vec p'_\alpha-\vec p'_\beta$, is always perpendicular to
$\vec p'_\alpha$ and never vanishes, so there is no strictly
stationary phase contribution to the integration over the
perpendicular component of $\vec r'$.
Furthermore, the magnitude of this momentum difference is roughly
independent of the perpendicular component of $\vec r'$, as it is
determined by the first part of these paths, which is identical for
$\alpha$ and $\beta$ (in the present example the first two scatterers
are identical).  However, the range of this integration, which we
denote by $l_\perp$, is finite --- roughly 10\% of the size of the
system, $L$, in the example --- and thus the action difference will
never grow beyond $l_\perp (\vec p'_\alpha-\vec p'_\beta)$ (both these
factors depend also on the parallel component of $\vec r'$, but their
product is roughly constant).  Now the question of whether this pair of
orbits contributes or not depends on the value of $\hbar$: if it is
much smaller than $10^{-3} p_F L$ there will be a fluctuating (never
stationary) phase factor, and the contribution will be unimportant,
while if $\hbar$ is much larger than $10^{-3} p_F L$ the phase will be
negligible throughout the whole integration range, and there will be a
finite contribution.

Quantitatively, such contributions may be described by a factor of
$h^{N-1} \, \overline{ \delta_\perp(\vec p'_\alpha-\vec p'_\beta) }$,
where the $\delta$ function is over the perpendicular components of
the momentum, and has a finite width
$\sim h/l_\perp$ and height $\sim l_\perp^{\, N-1}$.  In the analysis
of Secs.~III and IV the averaged distribution function
$\overline{ f_E }$ was introduced already at an earlier stage and the
integration over $\vec r'$ and $\vec p'$ was done in one step, so that
this $\overline{ \delta_\perp }$ function never appeared explicitly.
For long times there are exponentially many classical orbits from
$\vec r$ to $\vec r'$, with initial and final momenta within
$h/l_\perp$ of $\vec p$ and $\vec p'$ respectively, and thus the
averaging in $\overline{ f_E(\vec r',\vec p',t;\vec r,\vec p) }$ gives a
smooth distribution.  The fact that $\overline{f_E}$ is smooth makes it
invariant to the details of the averaging procedure.  Thus there is no
need to keep track of the actual distribution of sizes of $l_\perp$
(at least as long as the main contributions come from long orbits).

The crossover between `short' and `long' times for this purpose occurs
at an Ehrenfest time $t_E$, which depends on $\hbar$ and the typical
Lyapunov exponents.  A pair of paths starting very near to each other
in phase space will have their momentum difference multiplied by a
factor of order $l/R$ after each collision with an obstacle, where $l$
is the free path and $R$ is the radius of curvature of the obstacle.
Thus, if the initial momentum difference is $h/l_\perp$ and the
momentum difference at $t_E$ is required to be of order $p_F$, the
resulting estimate for the Ehrenfest time is
\beq
t_E \sim \tau { \log( p_F l_\perp /h) \over \log(l/R) }
\eeq
(suitably averaged values of $\tau$, $l$ and $l_\perp$ should be used here).
Often this estimate  can be taken to imply $t_E \sim \tau$, especially
in diffusive systems for which it is customary to take the limit of
small scatterers $R \ll l$ (recall that diffusive motion
is not a valid description for times $t \sim \tau$ anyway).  However,
in the semiclassical limit $\hbar$ will become so small that we will have
$t_E \gg \tau$, and in principle this should by taken into account.  For
instance, the magnetic field dependence is determined by the distribution of
areas enclosed by diffusive paths of length $t$, which should be replaced by
$t-2t_E$ since the paths are so close to each other that they enclose a
negligible area for a time $t_E$ near their beginning and end (this argument
does not apply to the contribution of periodic orbits).  Since the dominant
contributions come from long times, such corrections are unimportant
for diffusive systems, except for exceedingly small values of $\hbar$.
They are also unimportant in the chaotic systems studied here, because
of the assumption of ergodicity which was used for the individual
cavities (we have essentially assumed that both $t_E$ and the
time taken to traverse each cavity are much smaller than the
escape time from that cavity, which typifies the length of the
shortest relevant orbits).  It would be very interesting to study
systems for which $t_E$ is not negligible compared to the typical
propagation times.

The physics associated with the Ehrenfest time can be clarified by
considering a different geometry.  It is often claimed in the context
of disordered systems, on the basis of a perturbative analysis, that
any disordered system can be described by a single parameter --- the
diffusion constant --- on all length scales larger than some physical
cutoff such as the transport mean free path or the grain size for a
granular material.  However, an additional length--scale related to
the Ehrenfest time occurs naturally in the semiclassical analysis.
Furthermore, this length--scale depends not only on the microscopic
characteristics of the potential, but also on $\hbar$, and
diverges for $\hbar \rightarrow 0$.  As an example, consider a thick
slab of transparent disordered material.  If it is illuminated by a
{\it well--collimated} beam of light, a speckle pattern may be
observed in the transmitted light.  Based on the perturbative
analysis, one may expect that the condition on the thickness of the
slab is given by the transport mean--free--path, but in fact it is
clear that the relevant length is related to the Ehrenfest time
(provided that the microscopic features of the disordered sample are
`soft', i.e.\ they must be larger than a wavelength and must not act
as beam--splitters).  For slabs of thickness intermediate between
these two lengths, the beam will remain well collimated and no speckle
pattern will be observed, even though the direction of propagation of
the transmitted (or reflected) light will be random.

It is somewhat surprising that the importance of the averaging in
$\overline{ f_E(\vec r', \vec p',t;\vec r,\vec p) }$, or equivalently
the width of the $\overline{ \delta_\perp(\vec p'_\alpha \s- \vec p'_\beta)}$
functions, was not emphasized earlier.  This may be due in part to the
fact that much of the attention was devoted to physical quantities which
involve periodic classical orbits, such as the density of states
\cite{Gutz,LesH}.  In fact, this kind of averaging does not arise
naturally in the analysis of the contribution of periodic orbits
to the conductance either (see Sec.\ IV~C above), and even if it were
to occur, it would have no dramatic effect on integrals involving
$f_E(\vec r,\vec p,t;\vec r,\vec p)$.
In contrast, in the case of weak localization the averaging is
around the point $\overline{ f_E(\vec r,-\vec p,t;\vec r,\vec p) }$,
and it introduces qualitatively new types of orbits (i.e.\
non--self--retracing orbits) into the calculation for long times.
Likewise, the averaging for
$\overline{ f_E(\vec r',\vec p',t_a;\vec r,\vec p) } \;
 \overline{ f_E(\vec r',\vec p',t_b;\vec r,\vec p) }$ has an important
effect as it allows distinct classical orbits to overlap [see Fig.~3(a)].

\subsection{Corrections to the Conductance From Leading Order Propagators}

It is remarkable that the use of the Kubo formula, with a
self--consistent choice of the electric field, has enabled here the
calculation of corrections to the classical conductance which are
of higher order in $\hbar$, without having to evaluate such
high--order corrections to the propagators themselves.
As explained already in Sec.~II, the choice of the electric fields can
be shown to be unimportant on the basis of unitarity, and it is
somewhat surprising that the semiclassical results for
$\tensor \sigma(\vec r,\vec r')$ do not obey unitarity, order by order
in $\hbar$.  However, it turns out that the leading order results for
$\tensor \sigma(\vec r,\vec r')$
may give sub--leading contributions to the derivatives
$\nabla_{\vec r} \cdot \tensor \sigma(\vec r,\vec r')$.  In order to
clarify these issues, we briefly reconsider the example of a cavity with
ideally `random' scattering, which can be described by the circular
orthogonal ensemble (COE) of RMT \cite{RMT}.  As described in the following
paragraphs, this system can be observed to display all of
the above surprising aspects (i.e.\ a non-unitary leading order
approximation which still reproduces the quantum corrections to the
conductivity correctly, if the Kubo formula is used rather than the
Landauer formula), without invoking any semiclassical considerations.

Take the $n \times n$ scattering matrix, $S_{i,j}$, which is
associated with such a `random' cavity, connected to leads with a
total of $g_L \s+ g_R = n$ conducting channels.  The
dimensionless conductance is given in terms of this scattering matrix
by the Landauer formula
$g_C = \sum_{i=1}^{g_L} \sum_{j=g_L+1}^n |S_{i,j}|^2$.
Taking $S_{i,j}$ to be a random member of the COE, one may calculate
various averages of its matrix elements for any $n$.
For example, the fact that
$\aver{ |S_{i,j}|^2 } = (1 \s+ \delta_{i,j})/(n \s+ 1)$ implies that
$\aver{ g_C } = g_L g_R /(n \s+ 1)$.  Expansion in $1/n$ gives
$\aver{ g_C } \simeq g_L g_R (n^{-1} - n^{-2})$, in agreement with
the semiclassical results for an ergodic cavity,
Eqs.~(\ref{cl_fin}) and (\ref{WL_GC}).  Evaluation of the variance of
the conductance is slightly more complicated, although straightforward
\cite{RMT}:
\beqa
\aver{ (g_C)^2 }_{\text{Var}}  &\l=&
g_L g_R \aver{ (|S_{i,j}|^2)^2 }_{\text{Var}} +
g_L g_R (n \s- 2) \aver{ (|S_{i,j}|^2) (|S_{i,k}|^2) }_{\text{Var}}
\lf && \qquad\quad +
g_L g_R (g_L g_R \s- n \s+ 1)
\aver{ (|S_{i,j}|^2) (|S_{k,l}|^2) }_{\text{Var}}
\lf &=&
g_L g_R \LP {2 \over n(n \s+ 3)} - {1 \over (n \s+ 1)^2} \RP +
g_L g_R (n \s- 2) \LP {1 \over n(n \s+ 3)} - {1 \over (n \s+ 1)^2} \RP
\lf && \qquad\quad +
g_L g_R (g_L g_R \s- n \s+ 1)
\LP {n \s+ 2 \over n (n \s+ 1) (n \s+ 3)} - {1 \over (n \s+ 1)^2} \RP
\lf &\simeq& 2 g_L^2 g_R^2 / n^4  \; .
\eeqa
Here the indices $i,j,k,l$ which appear in the different type of
covariance terms are all taken to be different from each other. The
result is again in agreement with the semiclassical result of
\Eq{UCF_C} for an ergodic cavity.  Note that it is obtained due to a
cancelation of the first two contributions with each other, and that
knowledge of moments of $S_{i,j}$ to the second sub--leading
order in $1/n$ turns out to be necessary.  Obviously, if the leading
order expressions for the moments of $S_{i,j}$ were used in the Landauer
formula, only the classical conductance could have been calculated
correctly.

Consider now the freedom of using the known property of unitarity of
the scattering matrices, $\sum_j |S_{i,j}|^2 \s= 1$ for each row or column.
This allows us to replace  the contribution of each row of the
scattering matrix to the Landauer formula,
$\sum_{j=g_L+1}^{n} |S_{i,j}|^2$ by the combination
$(1 \s- \alpha) (1-\sum_{j=1}^{g_L} |S_{i,j}|^2) +
\alpha \sum_{j=g_L+1}^n |S_{i,j}|^2$, with any value of $\alpha$.
As the self--consistent
classical electric field for an ergodic cavity is concentrated in
the leads (cf.\ Sec.~II), it is possible to follow a line of derivation
equivalent to that used above for the SCA simply by choosing an
appropriate value of $\alpha$,
specifically $\alpha \s= g_L/n$.  Repeating this replacement for each
row and each column gives the ``Kubo formula'' for the conductance,
\beq \label{Kubo_C}
g_C = {g_L g_R \over n} - \sum_{i,j=1}^n E_i E_j |S_{i,j}|^2  \; ,
\eeq
with the ``classical electric field'' factors
\beq
E_i = \cases{g_R/n  & if $1 \leq i \leq g_L$ \cr
             -g_L/n & if $g_L \s+ 1 \leq i \leq n$ \cr}
\eeq
(the relative minus sign is due to the field being directed into the
cavity in one lead, and out of it in the other).
Observe that the classical conductance of the cavity is given here by a
``short--range part'', which does not require knowledge of the
``long--range'' scattering matrix at all.  It is stressed that
\Eq{Kubo_C} follows directly from the Landauer formula and the unitarity of
$S_{i,j}$.  In the present context it is considered to be more basic.
Evaluation of the mean quantum correction to the conductance can now
be performed using only the leading order result for the corresponding
moment of the matrix elements,
$\aver{ |S_{i,j}|^2 } \simeq (1 \s+ \delta_{i,j})/n$, despite the fact
that in this approximation the scattering matrix is not unitary.  Most
terms cancel with each other because $\sum_{i=1}^n E_i \s= 0$, and the
remaining $-\sum_{i=1}^n E_i^2/n$ gives the WL result.  The results
for UCF may likewise be obtained using only the leading order
expression for the moments,
$\aver{ (|S_{i,j}|^2) (|S_{k,l}|^2) }_{\text{Var}} \simeq
(\delta_{i,k} \delta_{j,l} \s+ \delta_{i,l} \delta_{j,k})/n^2$.
The analogy with the semiclassical calculation is thus complete.

It is expected that explicitly using the self--consistent electric
field could be very helpful in other calculation schemes too.  For
example, it should enable the calculations of Ref.~\cite{Ef} to be
done by expanding only to the second order in the diffuson--Cooperon
expansion, and not to the sixth order as was found necessary there
(this is based on counting the number of ladders in the different
diagrams for the ``long--range part'' of UCF \cite{KSL}).

In this system one can explicitly study the next order corrections to
the moments of $S_{i,j}$, and observe how their contributions could in
principle be as large as the WL and UCF terms being calculated, but
vanish due to cancellations between the different terms.  This
cancelation can be demonstrated without actually calculating the
higher order corrections, by arguing that $\sum_i E_i =0$ and using
the fact that the COE is insensitive to the actual values of the $i$ and
$j$ indices.  As noted in the closing paragraph of Sec.~II, a similar
cancelation can be demonstrated for diffusive systems using the
diagrammatic technique.  More generally, one may argue on the basis
of universality that additional contributions from higher order terms
should not be expected, at least in the cases of systems involving
ideally ergodic cavities (Sec.~V).  In order to verify this
expectation, it would be very interesting to study explicitly the
contribution of higher order corrections which are known for
semiclassical propagators \cite{Gasp}, especially those which are
known to involve $\hbar^{1/2}$ or $\hbar^{1/3}$ corrections, i.e.\ the
effects of diffraction and caustics.

In the field of disordered systems, higher order corrections in
$\hbar$ are usually analyzed in terms of two distinct small parameters
\cite{ArAlt}.  The first gives the accuracy of the
description of individual scattering events, and may be written as
$\lambda_F/l$ where $\lambda_F$ is the Fermi wavelength and $l$ is a
microscopic length such as the mean free path (for a semiclassical
analysis, an obvious microscopic length is the radius of curvature
of the obstacles depicted in Fig.~1).  It is often argued that
inaccuracy with respect to this parameter can be tolerated, because
the actual potential in a real system is unknown anyway, and instead
the mobility or the effective scattering cross--section are measured
directly.  The second small parameter is essentially
$\hbar/\Delta t_{esc} \sim 1/g$, where $\Delta$ is the mean
single--particle level spacing, $t_{esc}$ is the length of time a
typical electron spends in the sample before leaving through the leads,
and $g$ is the dimensionless conductance of the sample.  This
parameter describes the extent of lack of unitarity in the
semiclassical approach as discussed above.
Recently, it has been suggested that unitarity could be built into the
semiclassical approximation from the outset \cite{BK}.  It would be
very interesting to analyze this approach from the point of view of
two distinct small parameters.  Such an analysis could imply that the
SCA becomes, when unitarity is enforced, analogous to the nonlinear
$\sigma$--model --- the latter describes a disordered system with a
white--noise potential to all orders in $1/g$, but only to leading
order in $\lambda_F/l$.

\acknowledgements

The author wishes to thank H.~U. Baranger, Y.~Imry, R.~A. Jalabert,
A.~Kamenev, U.~Smilansky, A.~D. Stone, and D.~Ullmo for helpful discussions.
This work was supported by the German Israel Foundation (GIF) Jerusalem,
the Minerva Foundation (Munich, Germany), and the National Science
Foundation (under grants DMR93-08011 and PHY94-07194).

\figure{ Fig.~1:
Sketch of a path which together with its time reverse contributes to
weak localization.  The initial and final points may be varied along the
segment $l$.  The source (S) and drain (D) regions are connected through
ideal leads to particle reservoirs.
}

\figure{ Fig.~2:
Schematic sketch of conductance through a ballistic cavity, which is
considered to be completely chaotic and ergodic.  The electric field is
taken to be concentrated in regions of size $a$ in the left and right leads
(the results are independent of $a$).
}

\figure{ Fig.~3:
Sketch of the three different types of interference contributions to the
conductivity, together with some of the corresponding perturbative diagrams.
The classical paths $\alpha$ and $\beta$ start at $\vec r$ and end at
$\vec r'$.
(a) $\alpha$ different from $\beta$,
(b) $\alpha$ and $\beta$ lie on a periodic orbit $\gamma$, and one of them is
of negative duration,
(c) $\alpha$ and $\beta$ lie on $\gamma$, and $\beta$ is short (in this case
$\alpha$ traverses the whole of $\gamma$ and then repeats the segment $\beta$
a second time).  The sketch assumes diffusive motion with $t_E \sim \tau$.
}


\begin{references}
\bibitem{rems} For weak magnetic fields it is enough to take into
account the vector potential in the sensitive phase factors, and not
in the current operators.  As will become evident in Secs.~III and
IV, the symmetrization of the derivatives in $\hat \vec j(\vec r)$
is of no consequence as long as the SCA is being used.
\bibitem{KSLr} In a sense, this is the opposite of the approach taken in
Ref.~\cite{KSL}, in which the long--range part of the space--dependent
conductivity was included by convoluting each contribution with an
appropriately defined ``flow function''.  The point is that if integration
over the classical field configuration is performed, the flow function can
be replaced by a $\delta$ function.
\bibitem{remb} It is possible to use only backward directions, instead
of only forward directions --- this would arise naturally if we had
switched the roles of $\vec r$ and $\vec r'$, i.e.\
started from a description of the orbit $\alpha$ as following the path
$\beta$ first, and then continuing along the periodic orbit $\gamma$.
For the time--reversal symmetric systems considered here, it is
obvious that the results would not change.
\bibitem{repeti}
Any point $(\vec r,\vec p)$ on a periodic orbit $\gamma$
belongs in fact to many different periodic orbits, which differ
from each other by their periods $T_\gamma$.  These different periods
are integer multiples of the period of the `primitive' periodic orbit
which corresponds to $(\vec r,\vec p)$.  The orbits with different
repetition numbers differ from each other in their value of
$n_\gamma$, hence the necessity in principle for the time variable in
$n(\vec r, \vec p,t)$ (they share the same value of $D_\gamma$, for
example).  However, for any given time $t > t_E$ the overwhelming majority
of the periodic orbits are primitive, so that using $n(\vec r, \vec p)$,
with the smallest value of the corresponding $n_\gamma$ would
introduce only a negligible error.
\bibitem{comp} It is particularly impressive to compare the few lines
of algebra required in Sec.~V~B  with the
lengthy supersymmetric calculation of comparable quantities in
Ref.~\cite{IWZ}.
\end{references}
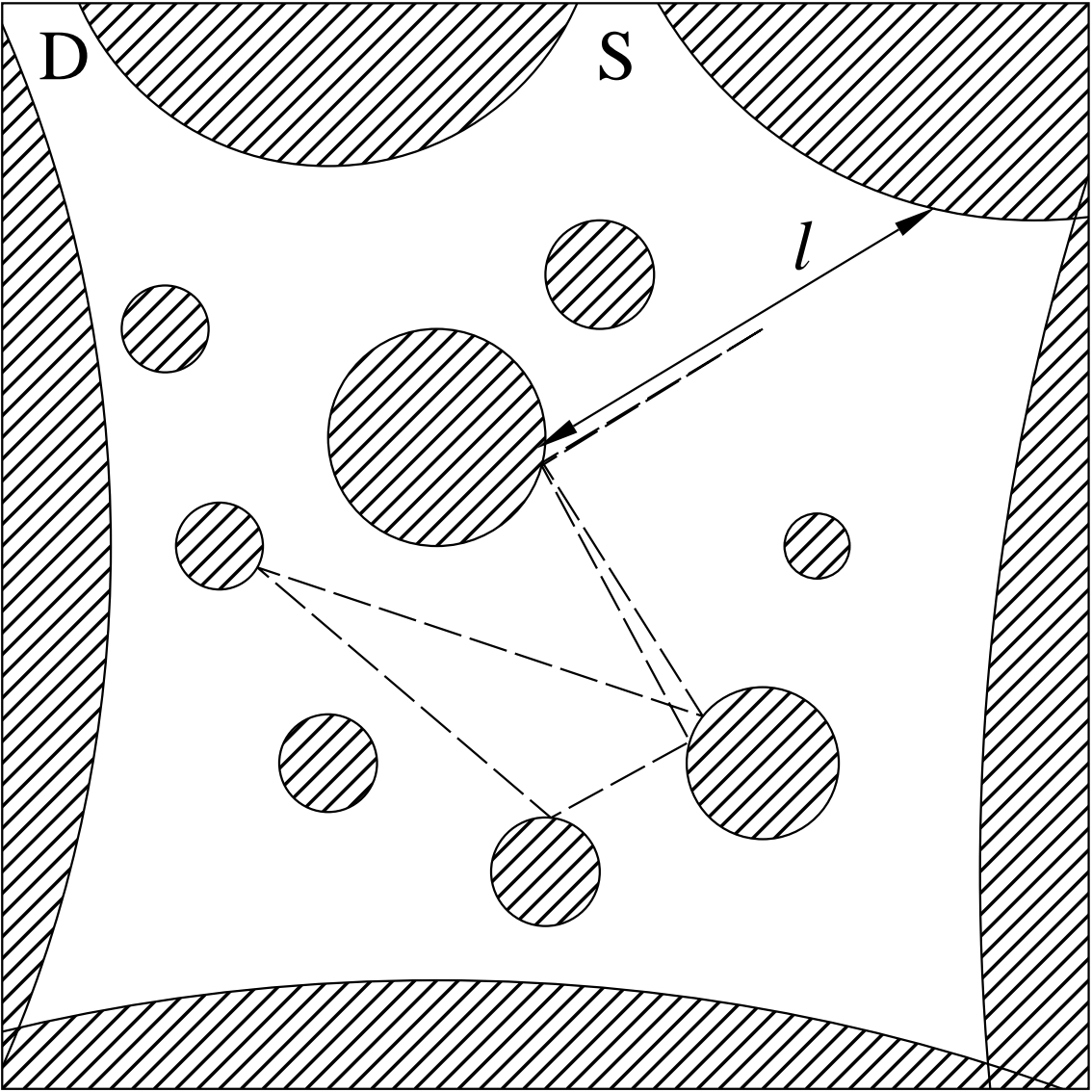
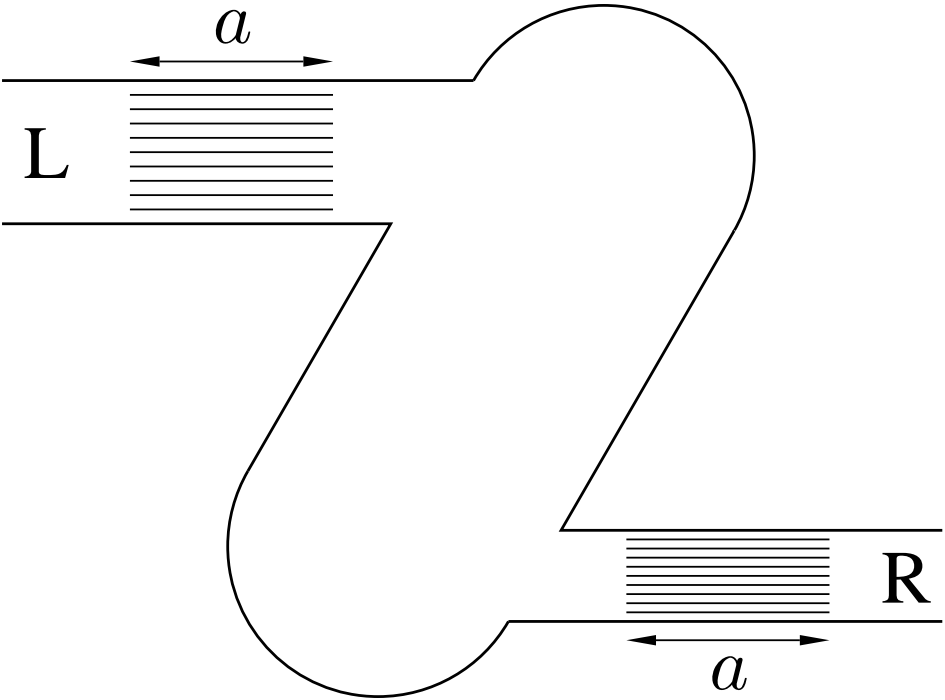
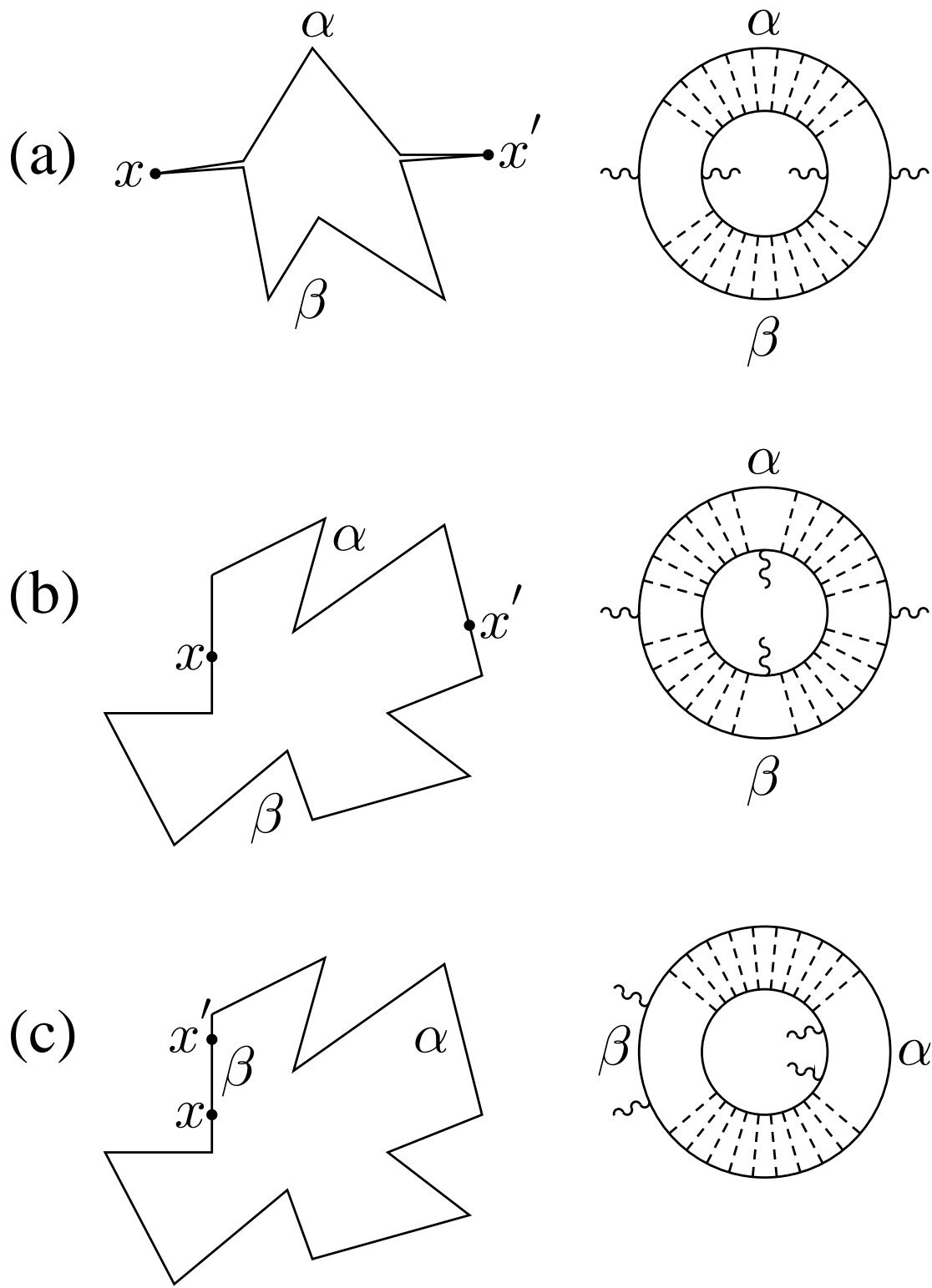
\end{document}